\documentclass{article}
\pdfoutput=1
\usepackage{float}
\usepackage{jheppub}
\usepackage{amsmath}
\usepackage{graphicx}
\usepackage{amssymb}
\usepackage{subfig}
\usepackage{natbib}
\usepackage[utf8]{inputenc}

\title{Jet-Images -- Deep Learning Edition}
\author{Luke de Oliveira,${}^a$}
\author{Michael Kagan,${}^{b}$}
\author{Lester Mackey,${}^c$}
\author{Benjamin Nachman,${}^{b}$ and}
\author{Ariel Schwartzman${}^b$}

\affiliation{$^{a}$ Institute for Computational and Mathematical Engineering, Stanford University, Stanford, CA 94305, USA}

\affiliation{$^{b}$SLAC National Accelerator Laboratory, Stanford University, 2575 Sand Hill Rd, Menlo Park,
  CA 94025, U.S.A.}

\affiliation{$^{a}$Department of Statistics, Stanford University, Stanford, CA 94305, USA}

\emailAdd{lukedeo@stanford.edu, mkagan@cern.ch, lmackey@stanford.edu, bnachman@cern.ch, sch@slac.stanford.edu}

\abstract{Building on the notion of a particle physics detector as a camera and the collimated streams of high energy particles, or jets, it measures as an image, we investigate the potential of machine learning techniques based on deep learning architectures to identify highly boosted $W$ bosons.  Modern deep learning algorithms trained on {\it jet images} can out-perform standard physically-motivated feature driven approaches to jet tagging.  We develop techniques for visualizing how these features are learned by the network and what additional information is used to improve performance. This interplay between physically-motivated feature driven tools and supervised learning algorithms is general and can be used to significantly increase the sensitivity to discover new particles and new forces, and gain a deeper understanding of the physics within jets.}

\begin{document}

\maketitle

\section{Introduction}
Collimated sprays of particles, called {\it jets}, resulting from the production of high energy quarks and gluons provide an important handle to search for signs of physics beyond the Standard Model (SM) at the Large Hadron Collider (LHC).  In many extensions of the SM, there are new, heavy particles that decay to heavy SM particles such as $W$, $Z$, and Higgs bosons as well as top quarks.  As is often the case, the mass of the SM particles is much smaller than the mass of the new particles and so they are imparted with a large Lorentz boost.  As a result, the SM particles from the boosted boson and top quark decays are highly collimated in the lab frame and may be captured by a single jet.  Classifying the origin of these jets and differentiating them from the overwhelming Quantum Chromodynamic (QCD) multijet background is a fundamental challenge for searches with jets at the LHC.  Jets from boosted bosons and top quarks have a rich internal substructure.   There is a wealth of literature addressing the topic of jet tagging by designing physics-inspired features to exploit the jet substructure (see e.g. Ref.~\cite{Altheimer:2012mn,Altheimer:2013yza,Adams:2015hiv}).  However, in this paper we address the challenge of jet tagging though the use of Machine Learning (ML) and Computer Vision (CV) techniques combined with low-level information, rather than directly using physics inspired features.  In doing so, we not only improve discrimination power, but also gain new insight into the underlying physical processes that provide discrimination power by extracting information learned by such ML algorithms. 

The analysis presented here is an extension of the jet-images approach, first introduced in Ref.~\cite{Cogan:2014oua} and then also studied with similar approaches by Ref.~\cite{Almeida:2015jua}, whereby jets are represented as images with the energy depositions of the particles within the jet serving as the pixel intensities.  When first introduced, jet image pre-processing techniques based on the underlying physics symmetries of the jets were combined with a linear Fisher discriminant to perform jet tagging and to study the learned discrimination information.  Here, we make use of modern deep neural networks (DNN) architectures, which have been found to outperform competing algorithms in CV tasks similar to jet tagging with jet images.  While such DNNs are significantly more complex than Fisher discriminants, they also provide the capability to learn rich high-level representations of jet images and to greatly enhance discrimination power.  By developing techniques to access this rich information, we can explore and understand what has been learned by the DNN and subsequently improve our understanding of the physics governing jet substructure.  We also re-examine the jet pre-processing techniques, to specifically analyze the impact of the pre-processing on the physical information contained within the jet.

Automatic feature extraction and high-level learned feature representations via deep learning have led to state-of-the-art performance in Computer Vision~\cite{vggnet,maxout:goodfellow,dropout:and:LRN}.  The focus of this work is on robust networks architectures to investigate what information and higher level representations a fully-connected multi-layer network and a convolutional neural network learn about jets.  There will be a focus on connecting the gains in performance with the underlying physical properties of jets through visualization. This paper is organized as follows:  The details of the simulated data sets and the definition of jet-images are described in Section~\ref{sec:simulation}.    The pre-processing techniques, including new insights into the relationship with underlying physics information, is discussed in Section~\ref{sec:preprocess}.  We then introduce the deep neural network architectures that we use in Section~\ref{sec:arch}.  The discrimination performance and the exploration of the information learned by the DNNs is presented in Section~\ref{sec:studies}.

\section{Simulation Details and the Jet Image}
\label{sec:simulation}

In order to study jet images in a realistic scenario, we use Monte Carlo (MC) simulations of high energy particle collisions. One important jet tagging application is the identification of highly Lorentz boosted $W$ bosons decaying into quarks amidst a large background from the generic production of quarks and gluons.  This classification task has been thoroughly studied experimentally\footnote{There is also an extensive literature on phenomenological studies - see references within the experimental papers.}~\cite{Khachatryan:2014vla,ATL-PHYS-PUB-2015-033,ATL-PHYS-PUB-2014-004} and used in many analyses~\cite{Aad:2015owa,Khachatryan:2014hpa,Khachatryan:2015mta,Khachatryan:2015oba,Khachatryan:2015gza,Khachatryan:2015bma,Khachatryan:2015cwa,Khachatryan:2015ywa,Aad:2014wea,Aad:2015agg,Aad:2015kna,Aad:2015ufa,Aad:2014haa}.  

To simulate highly boosted $W$ bosons, a hypothetical $W'$ boson is generated and forced to decay to a hadronically decaying $W$ boson ($W\rightarrow qq'$) and a $Z$ boson which decays invisibly ($Z\rightarrow \nu\bar{\nu}$).  The mass of the $W'$ boson determines the Lorentz boost of the $W$ boson in the lab frame since the $W'$ is produced nearly at rest and the $W$ boson momentum is approximately $m_{W'}/2$.  The invisible decay of the $Z$ boson ensures that the jet in the event with the highest transverse momentum is the $W$ boson jet.  Multijet production of quarks and gluons is simulated as a background.  Both the $W'$ signal and the multijet background are generated using \textsc{Pythia} 8.170~\cite{Pythia8,Pythia} at $\sqrt{s}=14$ TeV.  The minimum angular separation of the $W$ boson decay products in the plane transverse to the beam direction scales as $2m_{W}/p_{T,W}$, where $m_W\approx 80$~GeV and $p_{T,W}$ is the component of the $W$ boson momentum in this plane.  The tagging strategy and performance depend strongly on $p_{T,W}$, so we focus on a particular range: $250$~GeV~$<p_{T,W}<300$~GeV.  This corresponds to an angular spread of about $\Delta R = \sqrt{ \Delta \eta^2 + \Delta \phi^2} \sim 0.6$, where $\Delta \eta$ and $\Delta \phi$ are the distances between $W$ boson decay products in $(\eta, \phi)$ coordinates.  The decay products of the $W$ bosons as well as the background are clustered into jets using the anti-$k_t$ algorithm~\cite{antiktpaper} via \textsc{FastJet}~\cite{fastjet} 3.0.3.  To mitigate the contribution from the underlying event, jets are are trimmed~\cite{trimming} by re-clustering the constituents into $R=0.3$ $k_t$ subjets and dropping those which have $p_T^\text{subjet}<0.05\times p_T^\text{jet}$.  Trimming also reduces the impact of multiple proton-proton collisions occurring in the same event as the hard-scatter process (pileup).  We leave investgiation of the robustness of the neural network performance to pileup for future studies.

\begin{figure}[htbp!]
  \begin{center}
        \includegraphics[width=0.32\textwidth]{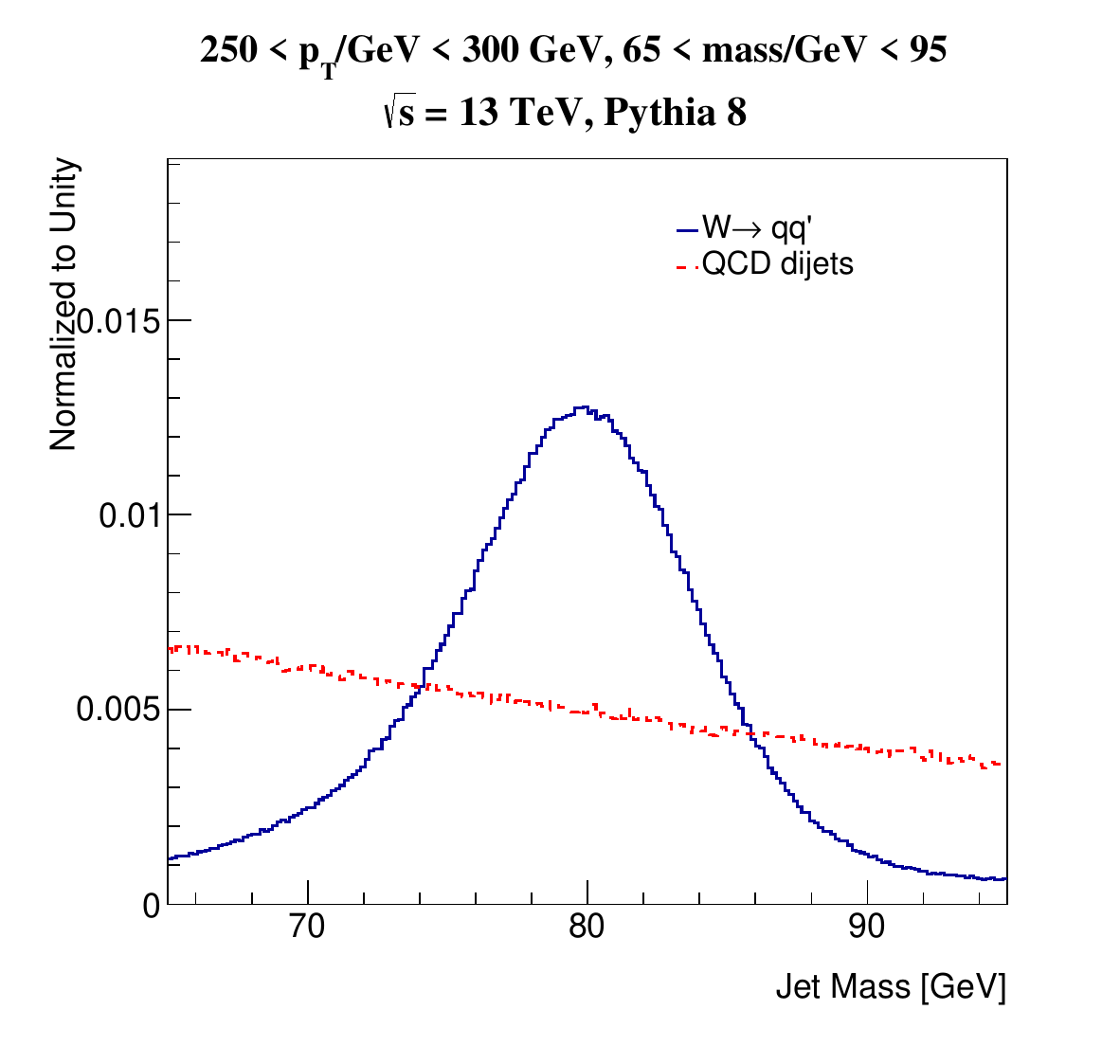} 
        \includegraphics[width=0.32\textwidth]{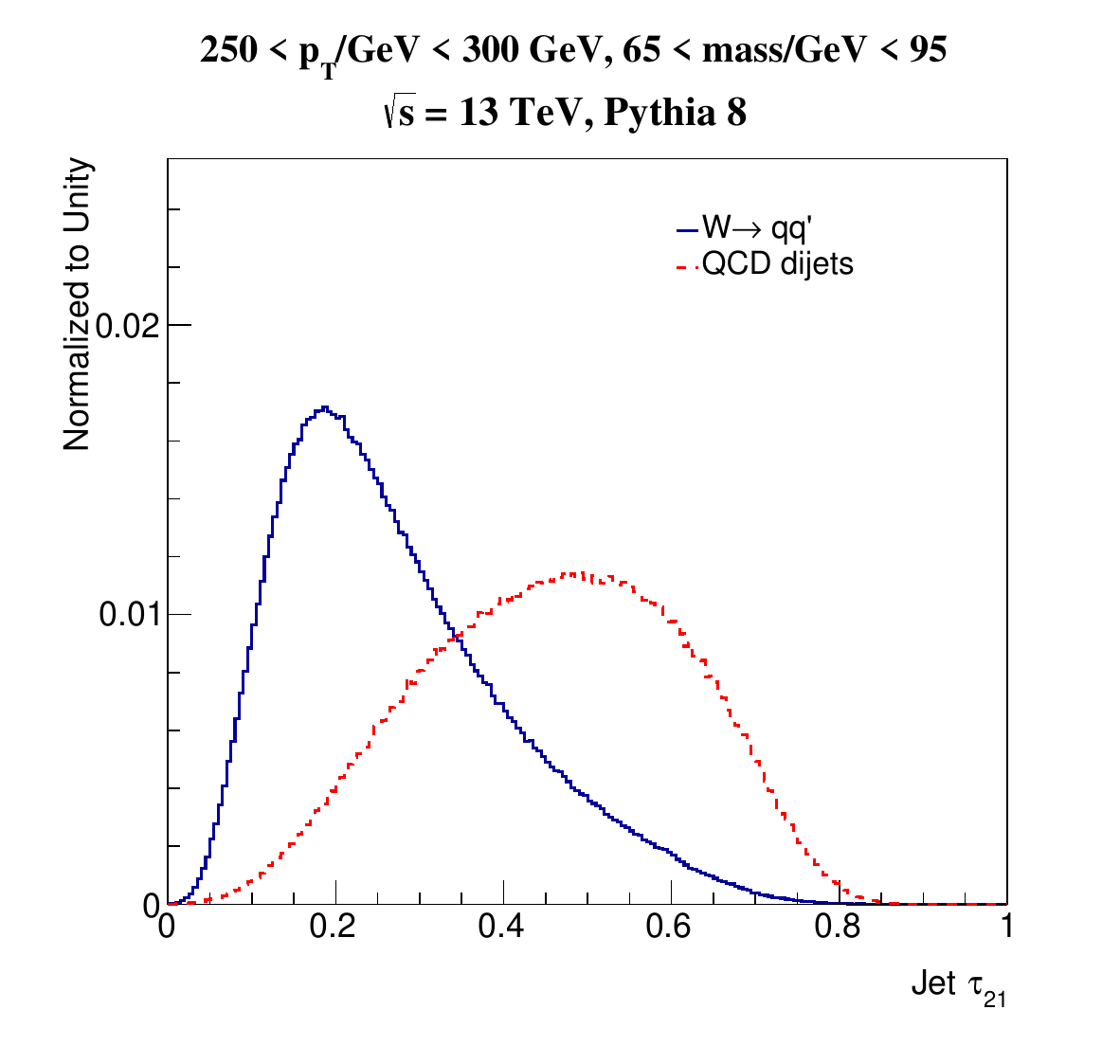} 
        \includegraphics[width=0.32\textwidth]{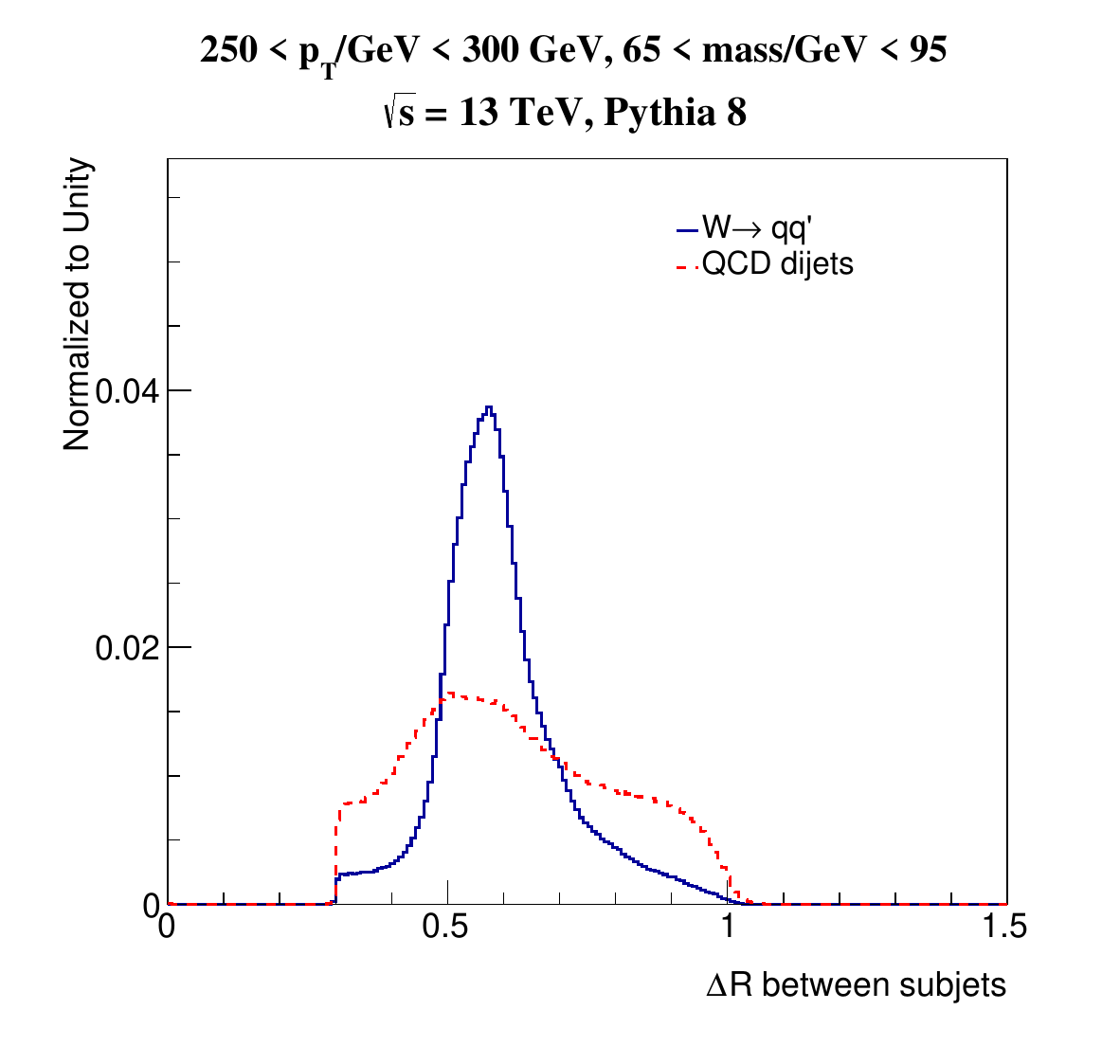}
      \caption{ The distributions of the jet mass (top left), $\tau_{21}$ (top right) and the $\Delta R$ between subjets (bottom) for signal (blue) and background (red) jets.
      \label{fig:datastats} }
    \end{center}
\end{figure}

Three key jet features for distinguishing between $W$ jets and QCD jets are the {\it jet mass}, {\it n-subjettiness}~\cite{nsub} and the distance in $(\eta,\phi)$ space between subjets of the trimmed jet ($\Delta R$).  The distributions of these three discriminating variables are shown in Fig.~\ref{fig:datastats}.   The jet mass is defined as $m_\text{jet}^2=\sum_{i,j} p_i p_j$, with jet constituent four-vectors $p_i$, and is a proxy for the boson mass in the case of $W$ boson events.  In the case of QCD background jets, the jet mass scales with the transverse momentum and the size of the jet.  $N$-subjettiness, in the form of $\tau_{21}$, is a measure of the likelihood that the jet has two hard prongs instead of one hard prong.  In this application, the winner-takes-all axis~\cite{Larkoski:2014uqa} is used to define the axis in the $\tau_{21}$ calculation.  One other useful feature is the jet transverse momentum.  However, since many of the other features have a strong dependence on the jet transverse momentum, we re-weight the signal so have the same $p_T$ distribution as the background.

To model the discretization and finite acceptance of a real detector, a calorimeter of towers with size $0.1\times 0.1$ in $(\eta,\phi)$ extends out to $\eta=5.0$.  The total energy of the simulated particles incident upon a particular cell are added as scalars and the four-vector $p_j$ of any particular tower $j$ is given by

\begin{align}
\label{eq:calo}
p_j = \sum_{i\text{ incident on $j$}}E_i(\cos\phi_j/\cosh \eta_j,\sin\phi_j/\cosh \eta_j,\sinh \eta_j/\cosh \eta_j,1),
\end{align}

\noindent where $E_i$ is the energy of particle $i$ and the center of the tower $j$ is $(\eta_j,\phi_j)$.  Towers are treated as massless.

 A {\it jet image} is formed by taking the constituents of a jet and discretizing its energy into pixels in ($\eta,\phi$), with the intensity of each pixel given by the sum of the energy of all constituents of the jet inside that ($\eta,\phi$) pixel.  We also investigate the use of the transverse projection of the energy in each tower as the pixel intensity.  In our studies, we take the jet image pixelation to match the simulated calorimeter tower granularity.  In the next section, we will discuss the nuances of standardizing the coordinates of a jet image as a pre-processing step prior to applying machine learning.  

\section{Pre-processing and the Symmetries of Space-time}
\label{sec:preprocess}

In order for the machine learning algorithms to most efficiently learn discriminating features between signal and background and to not learn the symmetries of space-time, the jet images are pre-processed.  This procedure can greatly improve performance and reduce the required size of the sample used for testing.  Our pre-processing procedure happens in four steps: translation, rotation, re-pixelation, and inversion.  To begin, the jet images are translated so that the leading subjet is at $(\eta,\phi)=(0,0)$.  Translations in $\phi$ are rotations around the $z$-axis and so the pixel intensity is unchanged by this operation.  On the other hand, translations in $\eta$ are {\it Lorentz boosts} along the $z$-axis, which do not preserve the pixel intensity.  Therefore, a proper translation in $\eta$ would modify the pixel intensity.  One simple modification of the jet image to circumvent this change is to replace the pixel intensity $E_i$ with the transverse energy $p_{T,i}=E_i/\cosh(\eta_i)$.  This new definition of intensity is invariant under translations in $\eta$ and is used exclusively for the rest of this paper\footnote{Transverse energy based pixel intensity was used in the original Jet-Images paper~\cite{Cogan:2014oua}}.

The second step of pre-processing is to rotate the images around the center of the jet.  If a jet has a second subjet, then the rotation is performed so that the second subjet is at $-\pi/2$.  If no second subjet exists, then the jet image is rotated so that the first principle component of the pixel intensity distribution is aligned along the vertical axis.  Unless the rotation is by an integer multiple of $\pi/4$, the rotated grid will not line up with the original grid.  Therefore, the energy in the rotated grid must be re-distributed amongst the pixels of the original image grid.  A cublic spline interpolation is used in this case - see Ref.~\cite{Cogan:2014oua} for details.  The last step is a parity flip so that the right side of the jet image has the highest sum pixel intensity.  

\begin{figure}[htbp!]
  \begin{center}
        \includegraphics[width=0.99\textwidth]{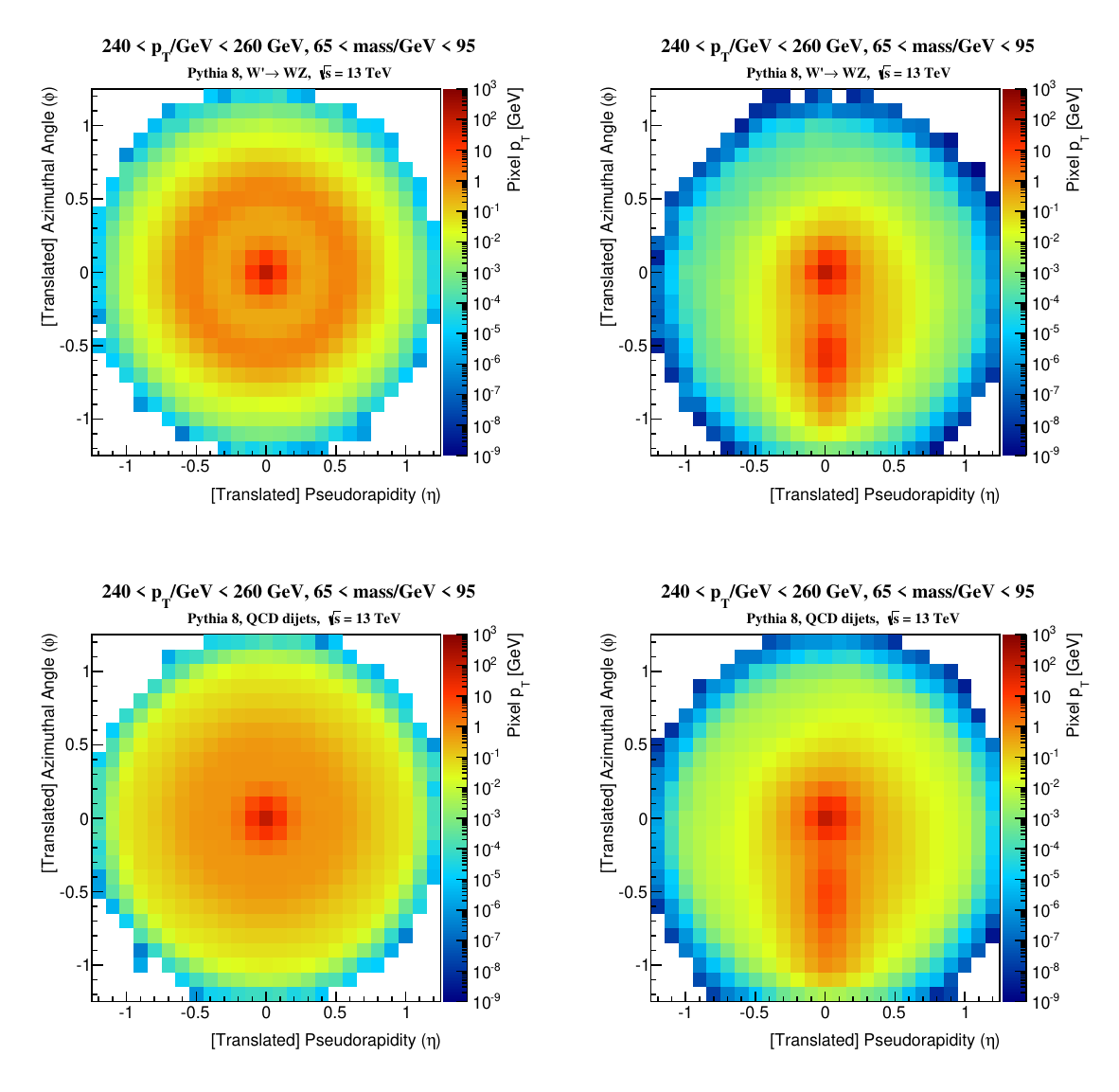}
      \caption{ The average jet image for signal $W$ jets (top) and background QCD jets (bottom) before (left) and after (right) applying the rotation, re-pixelation, and inversion steps of the pre-processing.  The average is taken over images of jets with $240$ GeV $<p_T<$ 260 GeV and 65~GeV~$<$ mass $<$~95~GeV.
      \label{fig:preprocess} }
    \end{center}
\end{figure}

Figure~\ref{fig:preprocess} shows the average jet image for $W$ boson jets and QCD jets before and after the rotation, re-pixelation, and parity flip steps of the pre-processing.  The more pronounced second-subjet can already be observed in the left plots of Fig.~\ref{fig:preprocess}, where there is a clear annulus for the signal $W$ jets which is nearly absent for the background QCD jets.  However, after the rotation, the second core of energy is well isolated and localized in the images.  The spread of energy around the leading subjet is more diffuse for the QCD background which consists largely of gluon jets, which have an octet radiation pattern, compared to the singlet radiation pattern of the $W$ jets, where the radiation is mostly restricted to the region between the two hard cores.

One standard pre-processing step that is often additionally applied in Computer Vision tasks is normalization.  A common normalization scheme is the $L^2$ norm such that $\sum I_i^2=1$ where $I_i$ is the intensity of pixel $i$.  This is particularly useful for the jet images where pixel intensities can span many orders of magnitude, and when there is large pixel intensity variations between images.  In this study, the jet transverse momenta are all around 250 GeV, but this can be spread amongst many pixels or concentrated in only a few. The $L^2$ norm helps mitigate the spread and thus makes training easier for the machine learning algorithm.  However, normalization can distort information contained within the jet image.  Some information, such as the Euclidean distance $\Delta R$ between subjets in $(\eta,\phi)$ is invariant under all of the pre-processing steps as well as normalization.  However, consider the {\it image mass}, 

\begin{align}
m_I^2=\sum_{i<j} E_iE_j(1-\cos(\theta_{ij})),
\end{align}

\noindent where $E_i=I_i/cosh(\eta_i)$ for pixel intensity $I_i$ and $\theta_{ij}$ is the angle between massless four-vectors with $\eta$ and $\phi$ at the $i$ and $j$ pixel centers.  The image mass is not invariant under all pre-processing steps but does encode key information to identify highly boosted bosons that would ideally be preserved by the pre-processing step.  As discussed earlier, with the proper choice of pixel intensity, translations preserve the image mass since it is a Lorentz invariant quantity.  However, the rotation pre-processing step does not preserve the image mass.  To understand this effect, consider two four-vectors: $p^\mu=(1,0,0,1)$ and $q^\mu=(0,1,0,1)$.   The invariant mass of these vectors is $\sqrt{2}$.  The vector $p^\mu$ is at the center of the jet image coordinates and the vector $q^\mu$ is located at $\pi/2$ degrees.  If we rotate the image around the jet axis so that the vector $q^\mu$ is at $0$ degrees, akin to rotating the jet image so that the sub-leading subjet goes from $\pi/2$ to $0$, then $p^\mu$ is unchanged but $q^\mu\rightarrow (1,0,\sinh(1),\cosh(1))$.  The new invariant mass of $q^\mu$ and $p^\mu$ is about 1, which is reduced from its original value of $\sqrt{2}$.  The parity inversion pre-processing step does not impact the image mass, but a $I^2$ normalization does modify the image mass.  The easiest way to see this is to take a series of images with exactly the same image mass but variable $I^2$ norm.  The map $I_i\mapsto I_i/\sum_j I_j^2$ modifies the mass by $m_I\mapsto m_I/\sum_j I_j^2$ and so the variation in the normalizations induces a smearing in the jet-image mass distribution.

\begin{figure}[htbp!]
  \begin{center}
        \includegraphics[width=0.5\textwidth]{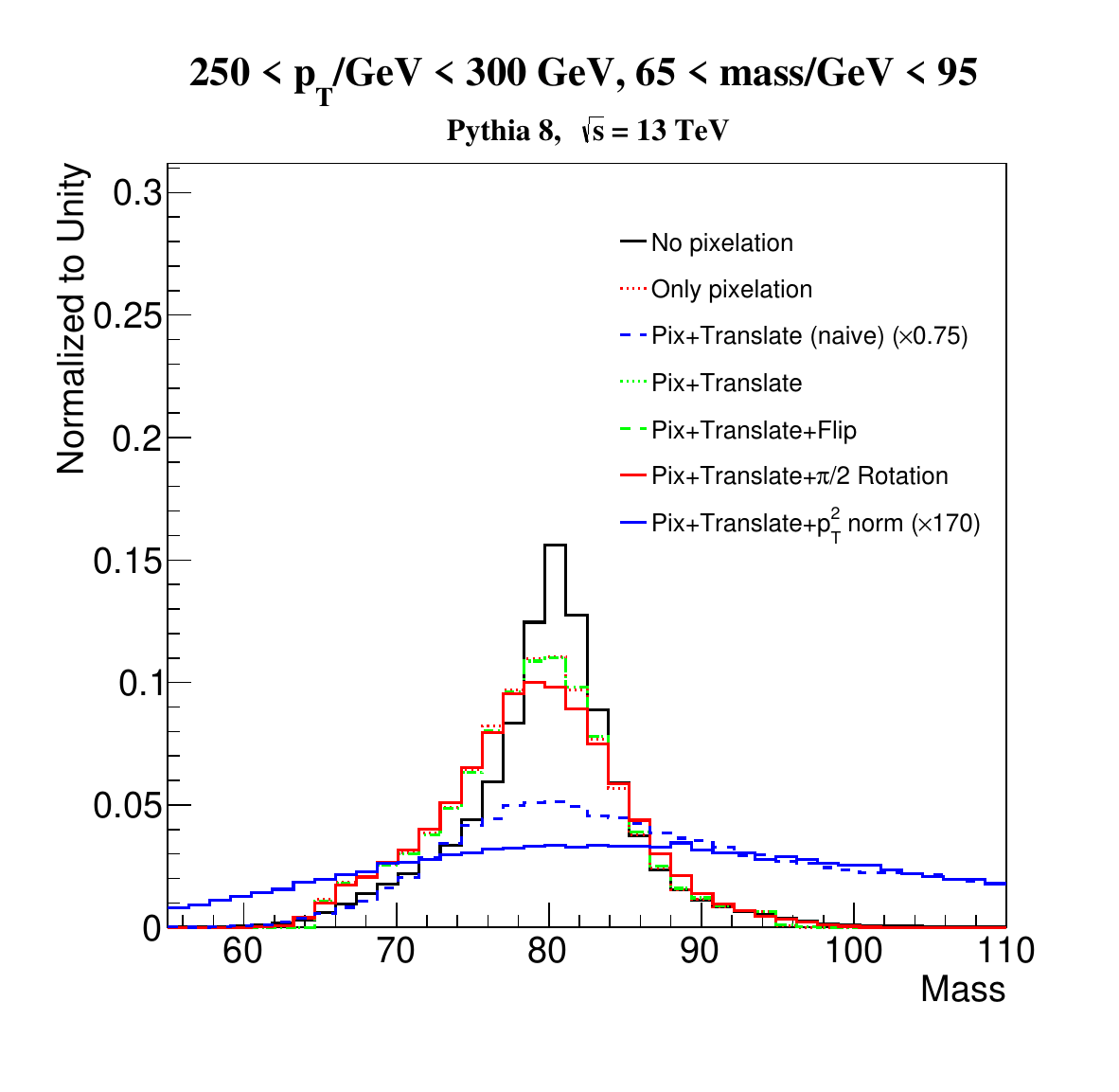}\includegraphics[width=0.5\textwidth]{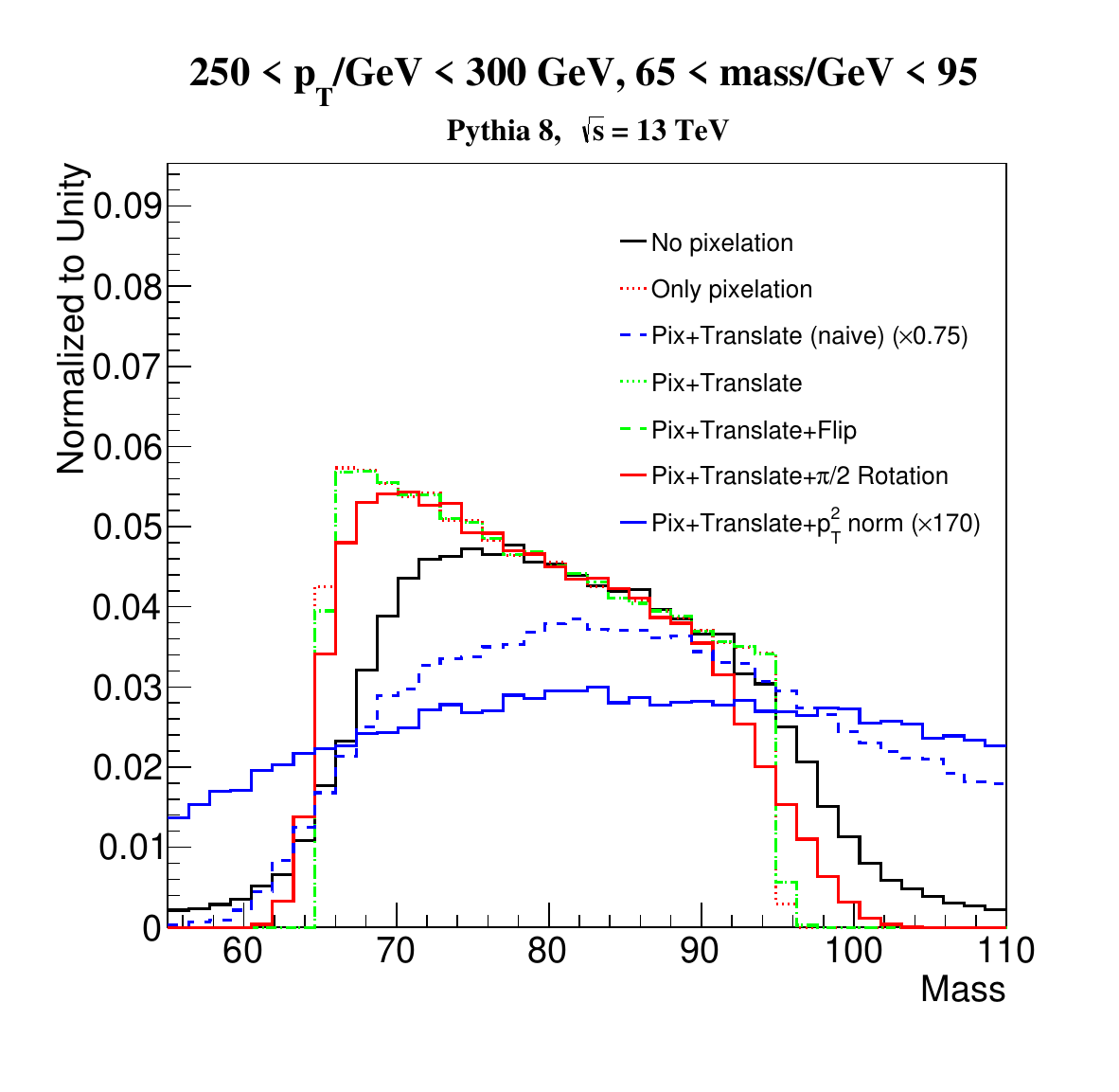}
      \caption{ The distribution of the image mass after various states of pre-processing for signal jets (left) and background jets (right).  The {\it No pixelation line} is the jet mass without any detector granularity and without any pre-processing.  {\it Only pixelation} has only detector granularity but no pre-processing and all subsequent lines have this pixelation applied as well as translation to center the image at the origin.  The translation is called {\it naive} when the energy is used as the pixel intensity instead of the pixel transverse momentum.  {\it Flip} denotes the parity inversion operation and the $p_T^2$ norm is a $L^2$ normalization scheme.  The naive translation and the $I^2$ normalization image masses are both multiplied by constants so that the centers of the distribution are roughly in the same location as for the other distributions.
      \label{fig:preprocess2} }
    \end{center}
\end{figure}

\begin{figure}[htbp!]
  \begin{center}
        \includegraphics[width=0.5\textwidth]{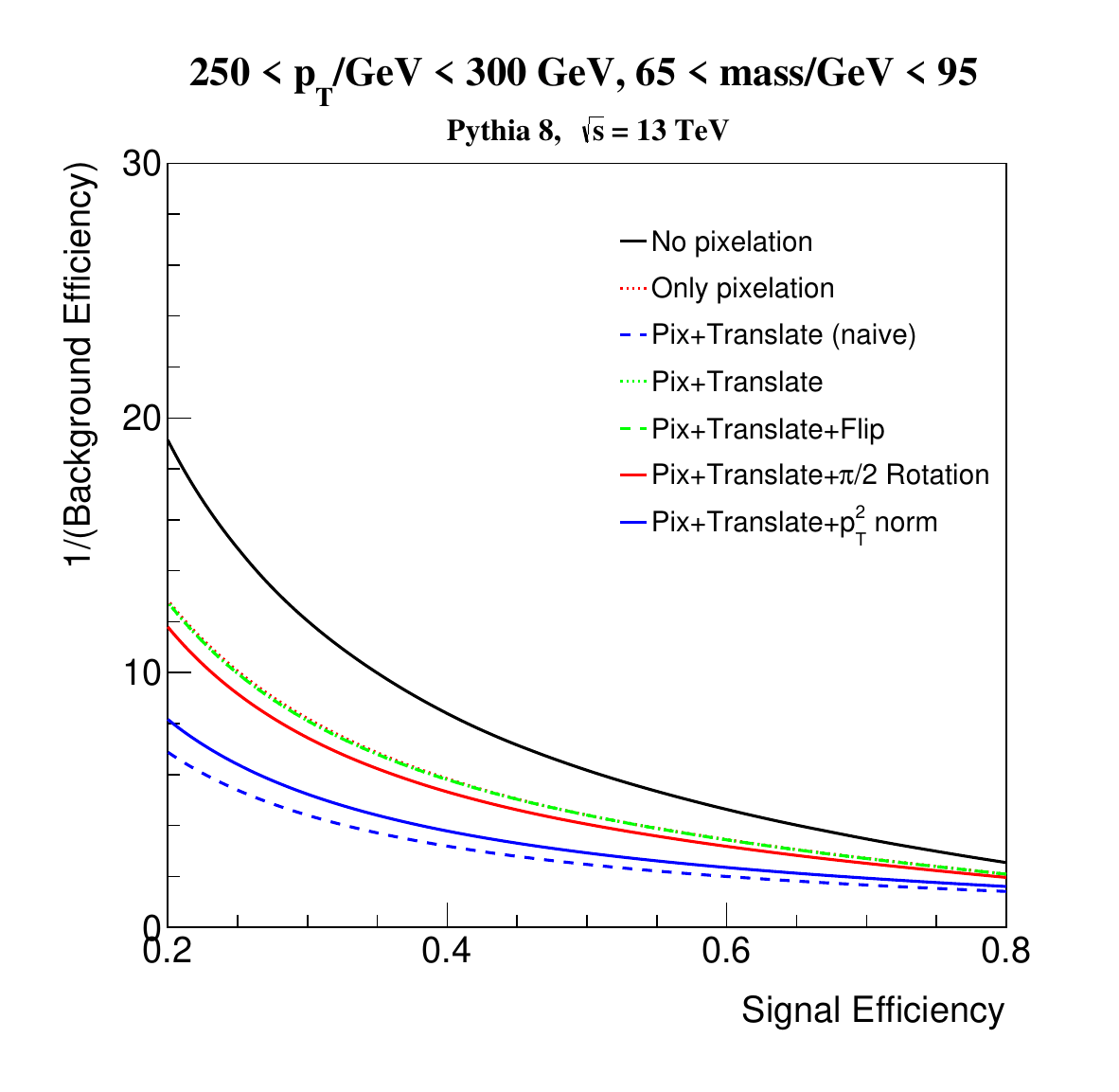}
      \caption{ The tradeoff between $W$ boson (signal) jet efficiency and inverse QCD (background) efficiency for various pre-processing algorithms applied to the jet (images).  The {\it No pixelation line} is the jet mass without any detector granularity and without any pre-processing.  {\it Only pixelation} has only detector granularity but no pre-processing and all subsequent lines have this pixelation applied as well as translation to center the image at the origin.  The translation is called {\it naive} when the energy is used as the pixel intensity instead of the pixel transverse momentum.  {\it Flip} denotes the parity inversion operation and the $p_T^2$ norm is a $L^2$ normalization scheme.  
      \label{fig:preprocess3} }
    \end{center}
\end{figure}

The impact of the various stages of pre-processing on the image mass are illustrated in Fig.~\ref{fig:preprocess2}.  The finite segmentation of the simulated detector slightly degrades the jet mass resolution, but the translation and parity inversion (flip) have no impact, by construction, on the jet mass.  The rotation that will have the biggest potential impact on the image mass is when the rotation angle is $\pi/2$ (maximally changing $\eta$ and $\phi$), which does lead to a small change in the mass distribution.  A translation in $\eta$ that uses the pixel energy as the intensity instead of the transverse momentum, which we refer to as a \textit{naive translation}, or an $L^2$ normalization scheme both significantly broaden the mass distribution.  One way to quantify the amount of information in the jet mass that is lost by various pre-processing steps is shown in the Receiver Operator Characteristic (ROC) curve of Fig.~\ref{fig:preprocess3}, which shows the inverse of the background efficiency versus the signal efficiency for passing a threshold on the signal-to-background likelihood ratio of the mass distribution (as described in Section~\ref{sec:studies}).  Information about the mass is lost when the ability to use the mass to differentiate signal and background is diminished.  The naive translation and the $I^2$ normalization schemes are significantly worse than the other image mass curves which are themselves similar in performance. 

\section{Network Architecture}
\label{sec:arch}

We begin with the notion that the discretization procedure outlined in Section \ref{sec:simulation} produces $25\times 25$ ``transverse-energy-scale'' images in one channel -- a High Energy Physics analogue of a grayscale image. We note that the images we work with are \emph{sparse} -- roughly 5-10\% of pixels are active on average (see Appendix~\ref{sec:sparsity} for details). Future work can build on efficient techniques for exploiting the sparse nature of these images. However, since speed is not our driving force in this work, we used convolution implementations defined for dense inputs.  We also study fully connected MaxOut networks~\cite{maxout:goodfellow}.  Other architectures were also studied, such as Stack Denoising Autoencoders~\cite{SDAE}, and multi-layer fully connected networks with various activation functions, but found that convolution and MaxOut networks were the most performant.

As a brief aside, we discuss some of the key neural network concepts which are used in the following section to describe our network architectures.  Fully connected (FC) layers take all features as input.  Convolution networks utilize convolution filters (or kernels) which are a set of weights $W$ that operate linearly on a small $n\times n$ (horizontal $\times$ vertical) patch of the input image.  For instance, a $3\times3$ filter takes as input a $3\times3$ patch of pixels and outputs $z = \sum_{i,j=1}^{3} x_{ij}W_{ij}$, where $x_{ij}$ is the input image patch.  The filter output can be considered as centered on that patch.  Each filter is convolved with the input image, in that the filter is applied to a given input patch and then moved horizontally and/or vertically to a new input patch on which the filter is applied.  By scanning over the entire image in this way, a the filter is convolved with the input, producing a convolved output. An important consideration when using convolutional networks is how one handles borders of images. Two main options exist -- one can consider only $n\times n$ patches that are fully contained within the input images, or one can consider every convolution that has at least one pixel from the image, \emph{zero-padding} as necessary to create valid convolutions. We use the latter, as we found better performance and better, more physics-driven filters. 

\begin{figure}[!htbp]
  \centering
  \includegraphics[width=0.75\textwidth]{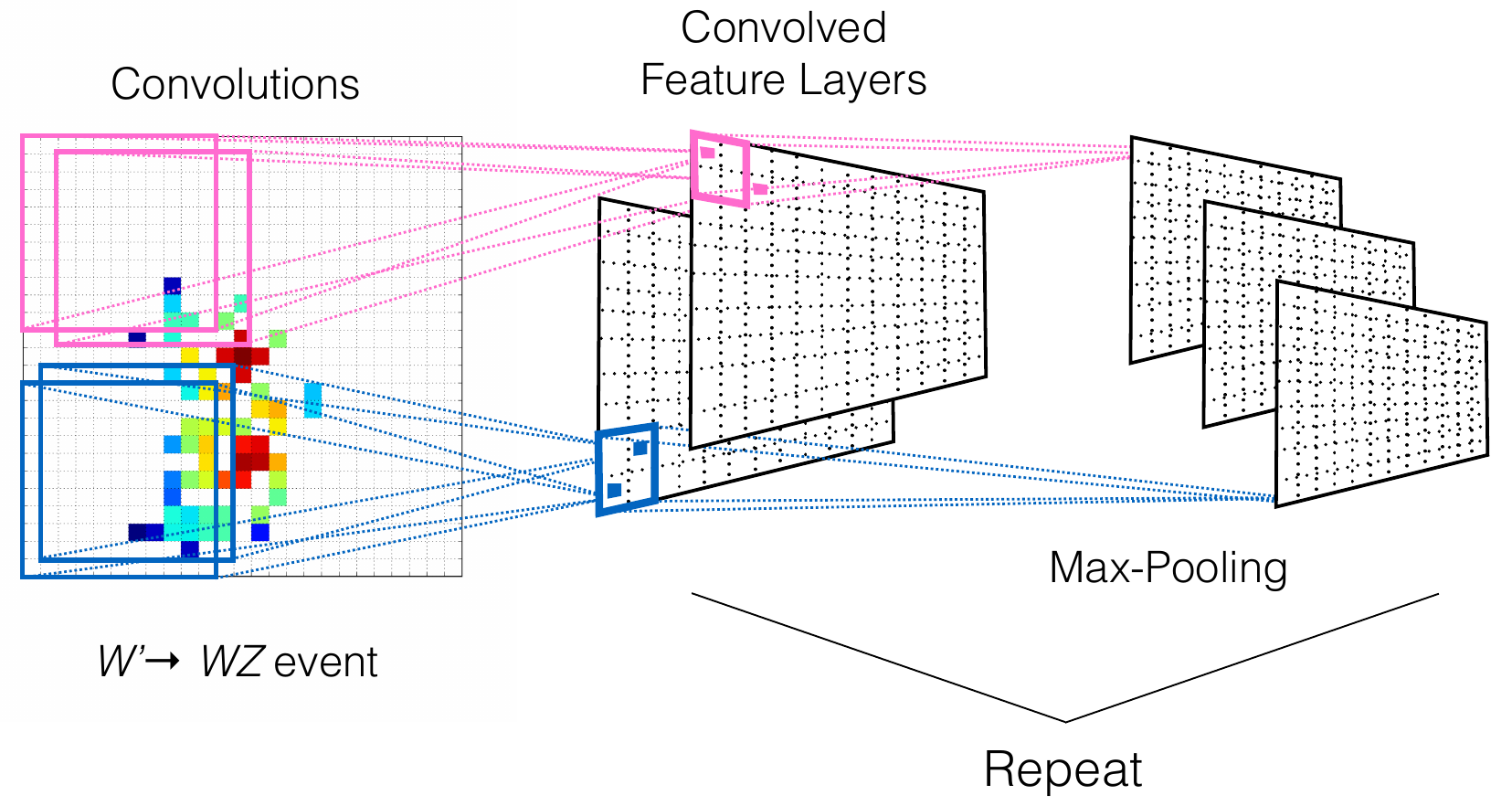}
  \caption{The convolution neural network concept as applied to jet-images.}
  \label{fig:arch}
\end{figure}

A non-linear activation function is typically applied to these convolution outputs, for which we use the Rectified Linear Unit (ReLU)~\cite{RELU} that takes an input $z$ and outputs $\max\{0,z\}$. ReLU's have been found to improve network training time, whilst having enough non-linear behavior to not degrade network performance. In addition, Rectified Linear Units do not suffer from a vanishing gradient, and speed up computation time while allowing for sparse networks by having true zero-valued activations.  After convolution(+activation) layers, a non-linear down-sampling is frequently performed using Max-pooling~\cite{MAXPOOL} which takes non-overlapping patches of convolution outputs as input, and outputs the maximum value for each patch.  A conceptual visualization of the convolution + Max-pooling network architecture that we employ can be seen in Figure~\ref{fig:arch}.

Finally, the MaxOut network makes use of the dense (Fully Connected) MaxOut activation unit, which takes an input vector $x$ and computes $k$ linear weightings $z_{j\in [1,k]} = \sum_{i} x_{i} W_{ij} + b_{j}$ and  outputs $\max_{j\in [1,k]}\ z_{j}$. Natural extensions of MaxOut layers to convolutional units exist, but were not examined. Conceptually, one can view the Rectified Linear Unit as a special case of the MaxOut with $k=2$ and with one of the weightings forced to output only zero. Though MaxOut units do not force sparsity of activation outputs in the same way as ReLU units, MaxOut networks provide the desirable attribute that they pair nicely with the model averaging effects of dropout in a natural way~\cite{maxout:goodfellow}.

\subsection{Architectural Selection} 
\label{ssub:architectural_selection}
For the MaxOut architecture, we utilize two FC layers with MaxOut activation (the first with 256 units, the second with 128 units, both of which have 5 piecewise components in the MaxOut-operation), followed by two FC layers with ReLU activations (the first with 64 units, the second with 25 units), followed by a FC sigmoid layer for classification. We found that the He-uniform initialization~\cite{HE_initialization} for the initial MaxOut layer weights was needed in order to train the network, which we suspect is due to the sparsity of the jet-image input. In cases where other initialization schemes were used, the networks often converged to very sub optimal solutions.  This network is trained (and evaluated) on un-normalized jet-images using the transverse energy for the pixel intensities

For the deep convolution networks, we use a convolutional architecture consisting of three sequential \texttt{[Conv + Max-Pool + Dropout]} units, followed by a local response normalization (LRN) layer~\cite{dropout:and:LRN}, followed by two fully connected, dense layers. We note that the convolutional layers used are so called ``full'' convolutions -- i.e., zero padding is added the the input pre-convolution. Our architecture can be succinctly written as:
\begin{equation}
  \mathtt{[Dropout \rightarrow Conv \rightarrow ReLU \rightarrow MaxPool] * 3 \rightarrow LRN \rightarrow [Dropout \rightarrow FC \rightarrow ReLU]  \rightarrow Dropout \rightarrow Sigmoid}.
\end{equation}

The convolution layers each utilize 32 feature maps, or filters, with filter sizes of $11\times 11$, $3\times 3$, and $3\times 3$ respectively.  All convolution layers are regularized with the $L^{2}$ weight matrix norm.  A down-sampling of $(2, 2)$, $(3, 3)$, and $(3, 3)$ is performed by the three max pooling layers, respectively.  A dropout~\cite{dropout:and:LRN} of 20\% is used before the first FC layer, and a dropout 10\% is used before the output layer.  The FC hidden layer consists of 64 units.

After early experiments with the standard $3\times 3$ filter size, we discovered significantly worse performance over a more basic MaxOut \cite{maxout:goodfellow} feedforward network. After further investigation into larger convolutional filter size, we discovered that larger-than-normal filters work well on our application. Though not common in the Deep Learning community, we hypothesize that this larger filter size is helpful when dealing with sparse structures in the input images. In Table~\ref{tab:kernelsize}, we compare different filter sizes,  finding the optimal filter size of $11\times11$, when considering the Area Under the ROC Curve (AUC) metric, based on the ROC curve outlined in Sections~\ref{sec:preprocess} and~\ref{sec:studies}.

\begin{table}[h!]
  \centering
  \begin{tabular}{l|c|c|c|c|c|c|c}
    \bfseries Kernel size &  $(3 \times 3)$ & $(4 \times 4)$ & $(5 \times 5)$ & $(7 \times 7)$ & $(9 \times 9)$  & $(11 \times 11)$ & $(15 \times 15)$ \\ 
    \hline
    \bfseries AUC & 14.770 & 12.452 & 11.061 & 13.308 & 17.291 & 20.286 & 18.140 
  \end{tabular}
  \caption{First layer convolution size vs. performance}
  \label{tab:kernelsize}
\end{table}

Two convolution networks, which differ in their pre-processing, are studied in this paper.  The first, which we refer to as the ConvNet, is trained (and evaluated) on un-normalized jet-images using the transverse energy for the pixel intensities.  The second, which we refer to as ConvNet-Norm, is trained (and evaluated) on $L^{2}$ normalized jet-images using the transverse-energy for the pixel intensities.  Examining the performance of both networks allows us to study the possible effects of normalization in the pre-processing.


\subsection{Implementation and Training} 
\label{ssub:implementation_and_training}

All Deep Learning experiments were conducted in Python with the Keras~\cite{Keras} Deep Learning library, utilizing NVIDIA C2070 graphics cards. One GPU was used per training, but several architectures were trained in parallel on different GPU's to optimize the performance of networks with different hyper-parameters.

We used 8 million training examples, with an additional 2 million validation samples for tuning the hyper-parameters, and 3 million testing samples.  Signal examples are weighted such that the total sum of weights is the same as the total number of background examples (as explained in Section~\ref{sec:simulation}).  These weights are used by the cost function in the training and in the ROC curve computations of the test samples.  The networks were trained with the Adam~\cite{DBLP:journals/corr/KingmaB14} algorithm (Stochastic Gradient Descent with Nesterov Momentum~\cite{Nesterov:1983wy} was also examined, but did not provide performance gains).  The training consisted of 100 epochs, with a 10 epoch patience parameter on the increase in AUC between 0.2 and 0.8 on a validation set.  Batch sizes of 32 were used for the MaxOut network, while batch sizes of 96 were used for the convolution networks.


\section{Analysis and Visualization} 
\label{sec:studies}


In this section, we examine the performance of the MaxOut and Convolution deep neural networks, described in Section~\ref{sec:arch}, in classifying boosted $W^\pm \to q q^\prime$ from QCD jets.  As one of our primary goals is to understand  what these NN's can learn about jet topology for discrimination, we focus on a restricted phase space of the mass and transverse momentum of the jets.  In particular, we restrict our studies to $250$ GeV $\leq p_T \leq 300$ GeV, and confine ourselves to a $65$ GeV $\leq m \leq 95$ GeV mass window that contains the peak of the $W$.   We also perform studies in which the discrimination power of the most discriminating physics variables has been removed, either though sample weighting or highly restrictive phase space selections, which allows us to focus on information learned by the networks beyond such known physics variables.  In this way, we construct a scaffolded and multi-approach methodology for understanding, visualizing, and validating neural networks within this jet-physics study, though these approaches could be used broadly.

The primary figure of merit used to compare the performance of different classifiers is the ROC curve.  The ROC curves allow us to examine the entire spectrum of trade-off between Type-I and Type-II errors\footnote{In this context, Type-I errors refer to incorrectly rejecting the signal, while Type-II errors refer to incorrectly accepting the background.}, as many applications of such classifiers will choose different points along the trade-off curve.   Since the classifier output distributions are not necessarily monotonic in the signal-to-background ratio, for each classifier we compute the  signal-to-background likelihood ratio\footnote{Practically, this is done by binning the distribution using variable width bins such that each bin has a fixed number of background events.  This number of background events is used to regulate the approximation and we check that the results are not sensitive to this choice.}.  The ROC curves are computed by applying a threshold to the classifier output likelihood ratio, and plotting the inverse of the fraction of background jet passing the threshold (the background rejection) versus the fraction of signal events passing the threshold (the signal efficiency).  We say that a classifier is \emph{strictly} more performant if the ROC curve is above a baseline for all efficiencies.  In decision theory, this is often referred to as domination (i.e. one classifier dominates another). It should be noted that any weights used to modify the distributions of jets (e.g. the $p_{T}$ weighting described in Section~\ref{sec:simulation}) are also used when computing the ROC curves.

%

For information exploration, several techniques were used:
\begin{itemize}

\item \textbf{ROC Curve Comparisons to Multi-Dimensional Likelihood Ratios:}  By combining several physics-inspired variables and computing their joint likelihood ratio, we can explore the difference between such multi-dimensional likelihood ratios and the neural networks' performance.  We also compute the joint likelihood ratio of the neural network output and physics-inspired variables.  If such joint classifiers improve upon the neural network performance, then we can consider the information in the physics-inspired variable (conditioned on the neural network output) as having been learned by the neural network.  If the joint classifier shows improved performance over the neural network, then the neural network has not completely learned the information contained in the physics-inspired variable.

\item \textbf{Convolution Filters:}  For convolution neural networks, we display the weights of the 11x11 filters as images.  These filters show how discrimination information is distributed throughout patches of the jets and give a view of the higher level representations learned by the network.  However, such filters are not always easy to interpret, and thus we also convolve each filter with a set of signal and background jet-images.  We then examine the difference between the  convolution output on the average signal jet-images and average background jet-images.  These difference give deeper insight into how the filters act on the jets to accentuate discriminating information.

\item \textbf{Joint and Conditional Distributions:}  We examine the joint and conditional distributions of various physics inspired features and the neutral network outputs.  If the conditional distribution of the physics variable $v$ given the neural network output $O$ is not independent of the neutral network output, i.e. $P(v|O) \neq P(v)\ \forall\ O$, then we consider the network to have learned information about this physics feature.

\item \textbf{Average, Difference, and Fisher Jet-Images:}  We examine average images for signal and background and their differences, as well as the Fisher Jets.  This is particularly illuminating when we select jets with specific values of highly discriminating physics-inspired variables.  This allows us to explore discriminating information contained in the jet images beyond the physics inspired variables.

\item \textbf{Neural Network Correlations per Pixel:}  We compute the linear correlations (i.e. Pearson correlation coefficient) between the neural network output and the distributions of intensity in each pixel.  This allows for a visualization of how the discriminating information learned by the neural network is distributed throughout the jet.  These visualizations are an approximation to the neural network discriminator and can be used to aid the development of new physics inspired variables (much like the Fisher Jet visualization).

\end{itemize}

\noindent The performance evaluation and information exploration techniques are examined in three settings, all of which require the aforementioned mass and transverse momentum selection.
\begin{enumerate}

\item \textbf{General Phase Space:} No alterations are made to the phase space.  This gives an overview of the performance and information learned by the networks

\item  \textbf{Uniform Phase Space:}  The weight of each jet is altered such that the joint distributions of mass, $n$-subjettiness, and $p_T$ are non-discriminative.  Specifically, we derive weights such that:
\begin{equation}
  f(m, \tau_{21}, p_T| W'\rightarrow WZ) \approx f(m, \tau_{21}, p_T| QCD).
\end{equation}
Both the weighting and network evaluation are performed in a slightly more restricted phase space requiring $\tau_{21}\in [0.2, 0.8]$. While $p_T$ is weighted in all phase space setting, mass and $n$-subjettiness are also weighted in this setting as they are amongst the most discriminating physics-inspired variables.  This weighting ensures that mass, $n$-subjettiness, and $p_T$ do not contribute to differences between signal and background, and thus this information is essentially removed from the discrimination power of the samples.  This allows us to examine what information beyond these variables has been learned and to understand where the neural network performance improvements beyond these physics derived variables comes from.  Neural networks that are trained in the General Phase Space are applied as the discriminant under this ``flattening'' transformation. We also use the training weights inside this window to train an additional convolution network. We look for increases in performance that would indicate information learned beyond the information contained in the weighted physics variables.

\item \textbf{Highly Restricted Phase Space:} The phase space of mass, $n$-subjettiness, and $p_T$ are restricted to very small windows of size: $m\in [79, 81]$ GeV,  $p_T \in [250, 255]$ GeV, and  $\tau_{21} \in [0.19, 0.21]$. No weighting (beyond the $p_{T}$ weighted described in Section~\ref{sec:simulation}) is performed, and the networks trained in the General Phase Space are used for discrimination and evaluation.  This highly restricted window provides a different method to effectively remove the discrimination power of mass, $n$-subjettiness, and $p_T$ as there is little to no variation of the variables in this phase space for either signal or background.  Thus, any discrimination improvements of the neural networks over the physics-inspired variables would be coming from information learned beyond these variables.  While the weighting in the Uniform  Phase Space is designed also to remove such discrimination, it produces a non-physical phase space.  The Highly Restricted Phase Space allows us to ensure that the neural network performance improvements are valid and transferrable to a less contrived phase space.

\end{enumerate}
By examining the performance of the neural networks in these different phase spaces, we aim to systematically remove known discriminative information from the networks' performance and thereby probe the information learned beyond what is already known by physics inspired variables.


\subsection{Studies in the General Phase Space} 
\label{sub:coarse_studies}

\begin{figure}[!htbp]
\begin{center}
\subfloat[]{
	\includegraphics[width=0.48\textwidth,angle=0]{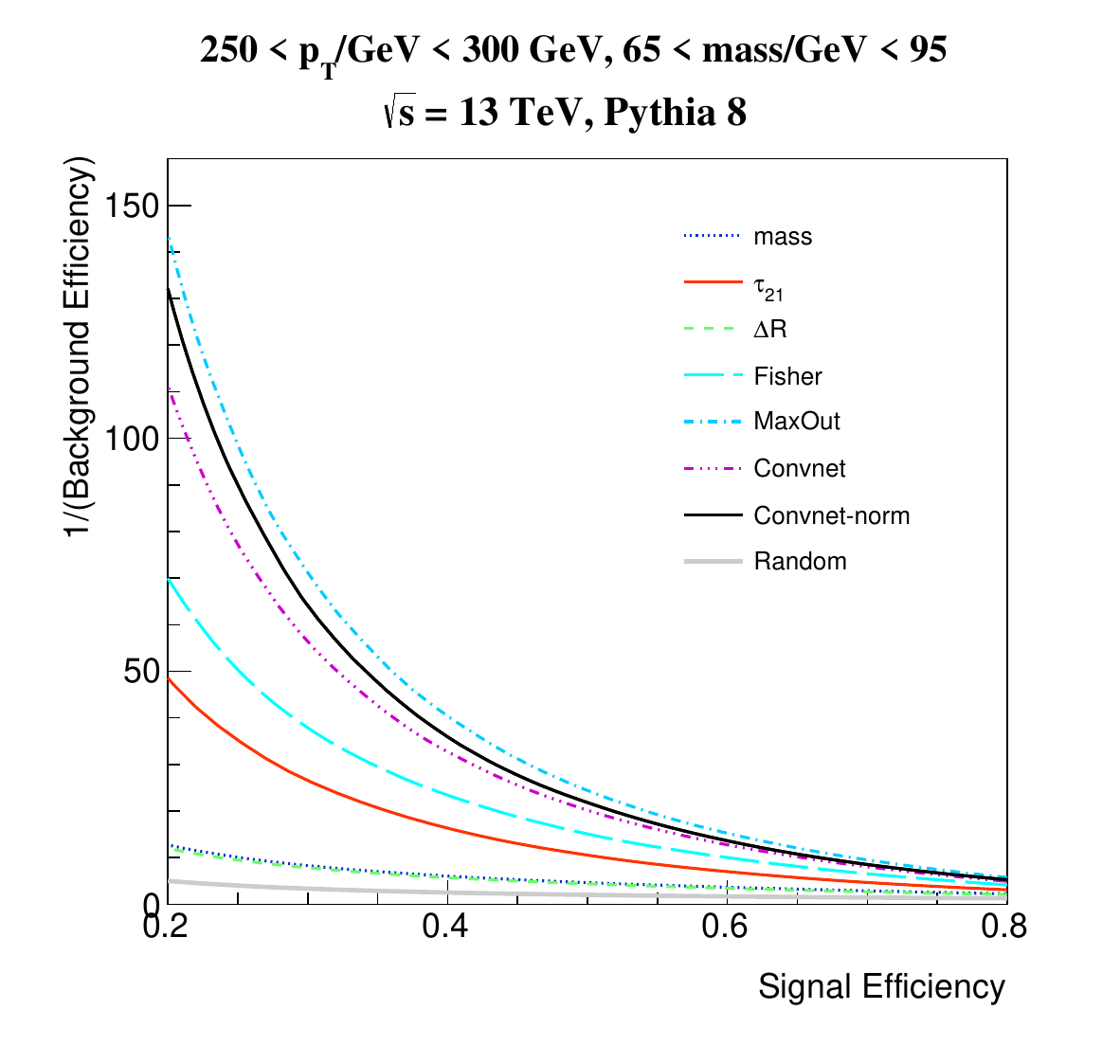}
	\label{fig:combinedROC1a}
}
\subfloat[]{
	\includegraphics[width=0.48\textwidth,angle=0]{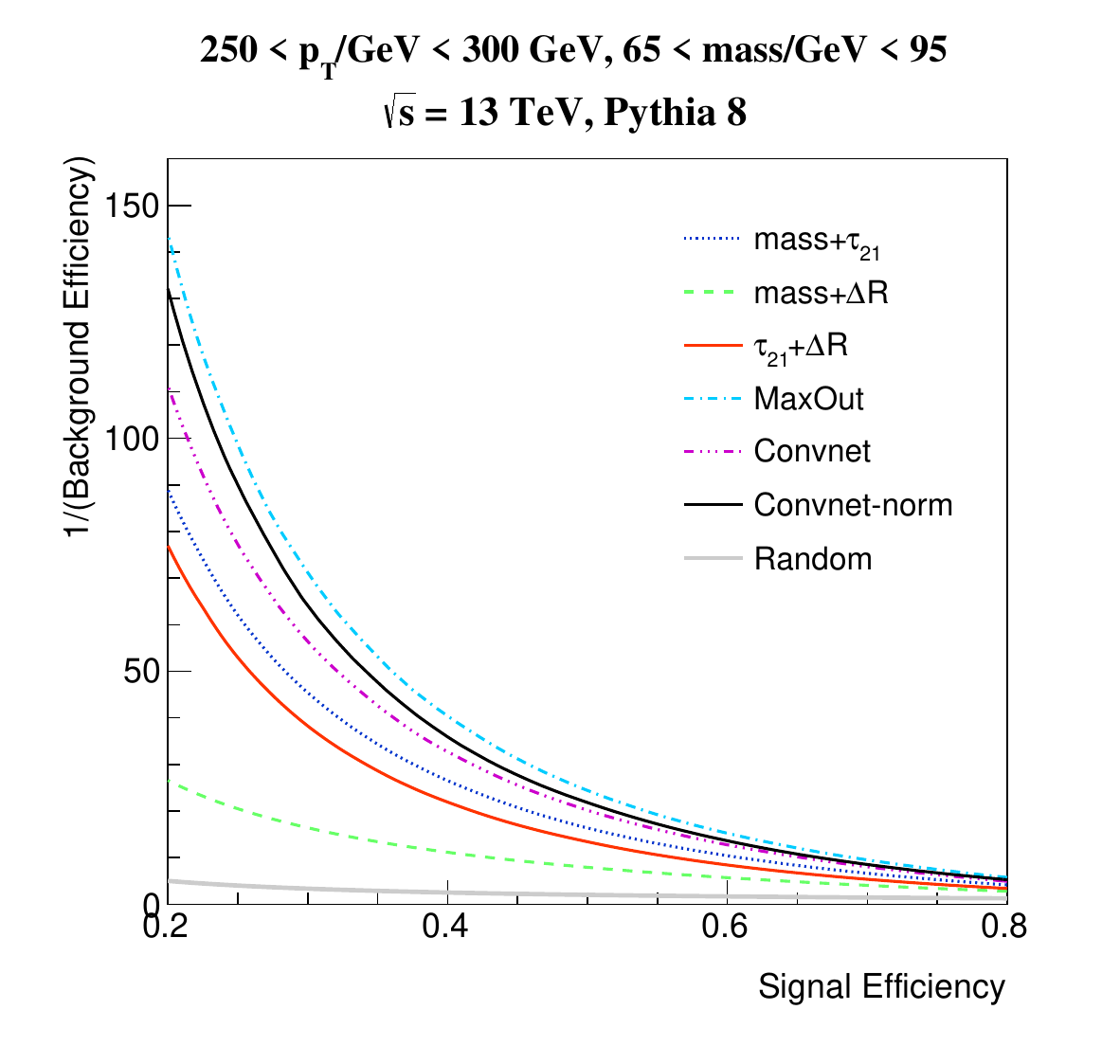}
	\label{fig:combinedROC1b}
}
\end{center}
   \caption{Left: ROC curves for individual physics-motivated features as well as three deep neural network discriminants.  Right: the DNNs are compared with pairwise combinations of the physics-motivated features.}
  \label{fig:combinedROC1}
\end{figure}

In order to evaluate the overall discrimination performance of the DNNs to that of the physics-driven variables, we examine the ROC curves in Figure~\ref{fig:combinedROC1}. In particular, we compare the DNNs to $n$-subjettiness~\cite{nsub} $\tau_{21} = \tau_{2}/\tau_{1}$, the jet mass, and the distance $\Delta R$ between the two leading $p_{T}$ subjets.  In Figure~\ref{fig:combinedROC1a}, we can see that the three DNNs have similar performance, but the MaxOut networks outperforms the ConvNet networks.  We suspect that the MaxOut outperforms the ConvNets due to sparsity of the jet-images, whereby the MaxOut network views the full jet-image from the inital hidden layer while the sparsity tends to make it difficult for the ConvNets to learn meaningful convolution filters.  We also see that the ConvNet-Norm outperforms the ConvNet trained on the un-normalized jet-images.  We observe that the classification performance of the ConvNet discriminant is highest when jet images are normalized, despite the fact that image normalization destroys jet mass information from the images. As we will see soon, it is difficult for these networks to fully learn the jet mass, so the lack of of mass information from pre-processing does not necessarily lead to worse discrimination performance. On the other hand, normalization is having an impact on the ability to effectively train the ConvNet network on jet images.  Finally, we see that the DNNs significantly improve the discrimination power relative to the Fisher-Jet discriminant\footnote{The Fisher discriminant is trained in three partitions of $\Delta  R$ ($\Delta R \in [0.25, 0.5],\ [0.5, 0.75],\ [>0.75])$, in order to account for the non-linear variation in jet-images from the differing positions of the two subjets.  Also note that unlike in the original implementation, here we do not normalize the jet images when computing the Fisher Jet.  This leads to slightly better performance.}, as described in reference~\cite{Cogan:2014oua}. In addition, in Figure~\ref{fig:combinedROC1b} we see that the DNNs also outperform the two-variable combinations of the physics inspired variables (computed using the 2D likelihood ratio\footnote{This is computed using the same regulated binning scheme as the 1D likelihoods described earlier.}).   It is interesting to note that combining mass and $\tau_{21}$, or $\tau_{21}$ and $\Delta R$, achieve much higher performance than the individual variables and are significantly closer to the performance of the DNNs.  However, the large difference in performance between the DNNs and the physics-variable combinations implies the DNNs are learning information beyond these physics variables.

\begin{figure}[!htbp]
  \begin{center}
\subfloat[]{
	\includegraphics[width=0.48\textwidth,angle=0]{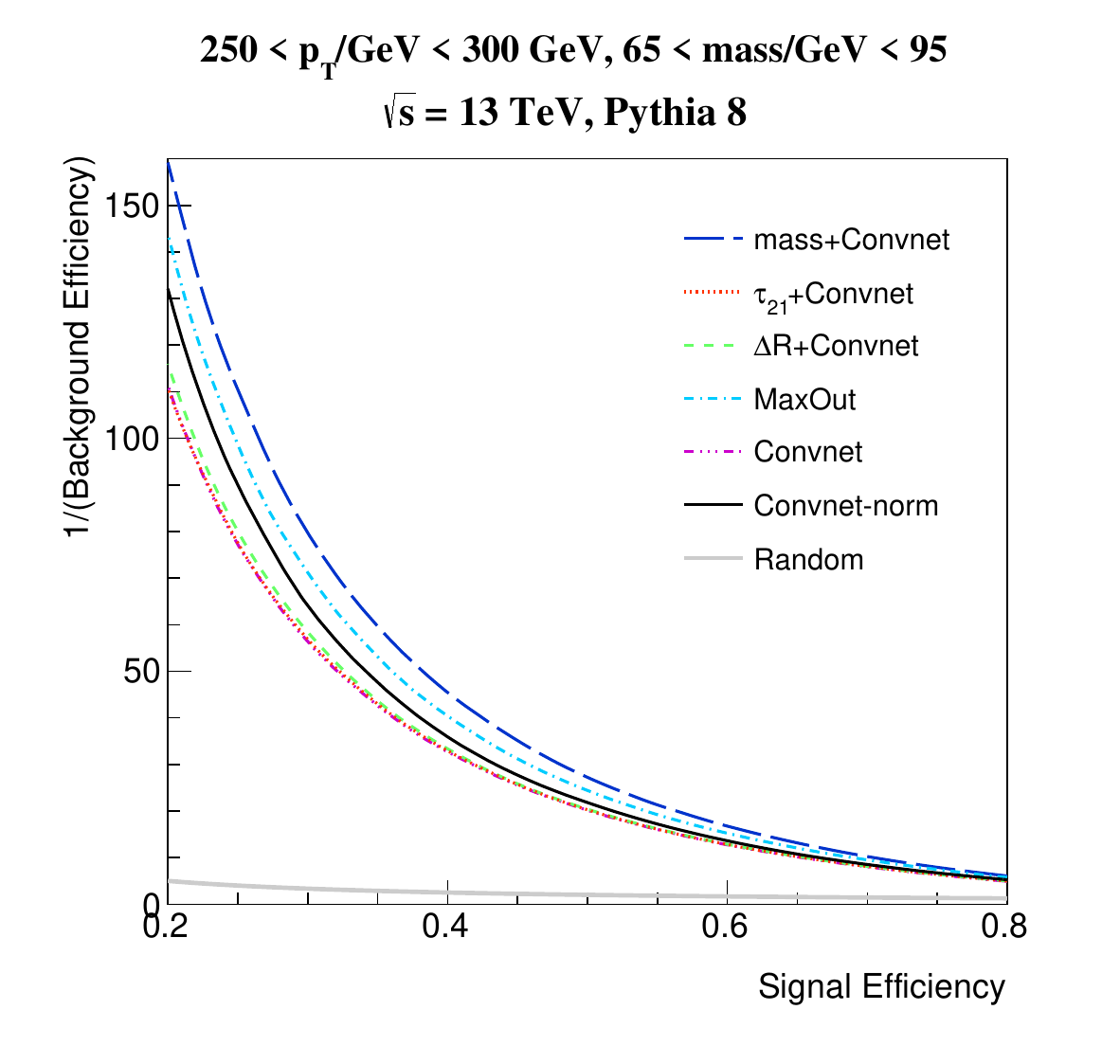}
	\label{fig:combinedROC2a}
}
\subfloat[]{
	\includegraphics[width=0.48\textwidth,angle=0]{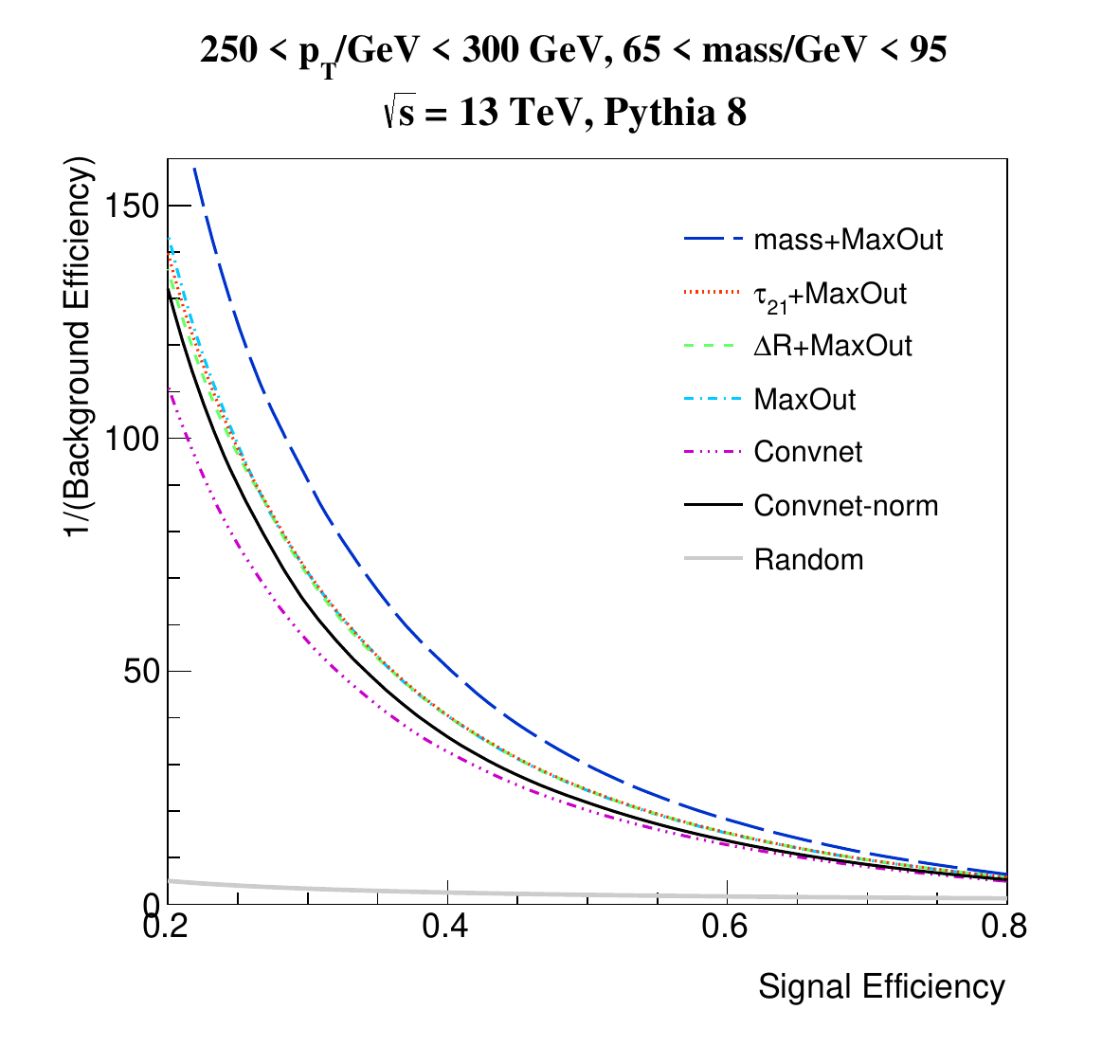}
	\label{fig:combinedROC2b}
}
\end{center}
  \caption{ROC curves that combined the DNN outputs with physics motivated features for the Convnet (left) and MaxOut (right) architectures.}
  \label{fig:combinedROC2}
\end{figure}

While we can see in Figure~\ref{fig:combinedROC1} that the DNNs outperform the individual and two-variable physics inspired discriminators, we want to understand if these physics variables have been learned by the networks.  As such, we compute the combination of the DNNs with each of the physics inspired variables (using the 2D likelihood), as seen for the ConvNet in Figure~\ref{fig:combinedROC2a} and for the MaxOut network in Figure~\ref{fig:combinedROC2b}. In both cases, we see that the discriminators combining $\Delta R$ or $\tau_{21}$ with the DNNs does not improve performance.  This indicate that the discriminating information in these variables relevant for the classification task has already been fully learned by the networks\footnote{This is not strictly speaking true, since there may be other variables that are needed in order to fully capture the full information of a given variable.  For example, consider independent random variables $X_i$ that are $\pm 1$ with probability $1/2$.  If $Y=X_1X_2$, then $X_1$ is independent of $Y$ but the joint distribution of $(X_1,X_2)$ is not independent of $Y$.  The statement is true in the absence of interactions with other variables.}.  However, adding mass in combination with the DNNs shows a noticeable improvement in performance over the DNNs alone.  This indicates that not all of the discriminating information relevant for jet tagging contained in the mass variable has been learned by the DNNs.  While it is not shown, similar patterns are found for the Convnet-Norm network.

\begin{figure}[htbp!]
  \begin{center}
  
  \subfloat[ ConvNet\label{fig:sculptedConv}]
      {
        \includegraphics[width=0.32\textwidth]{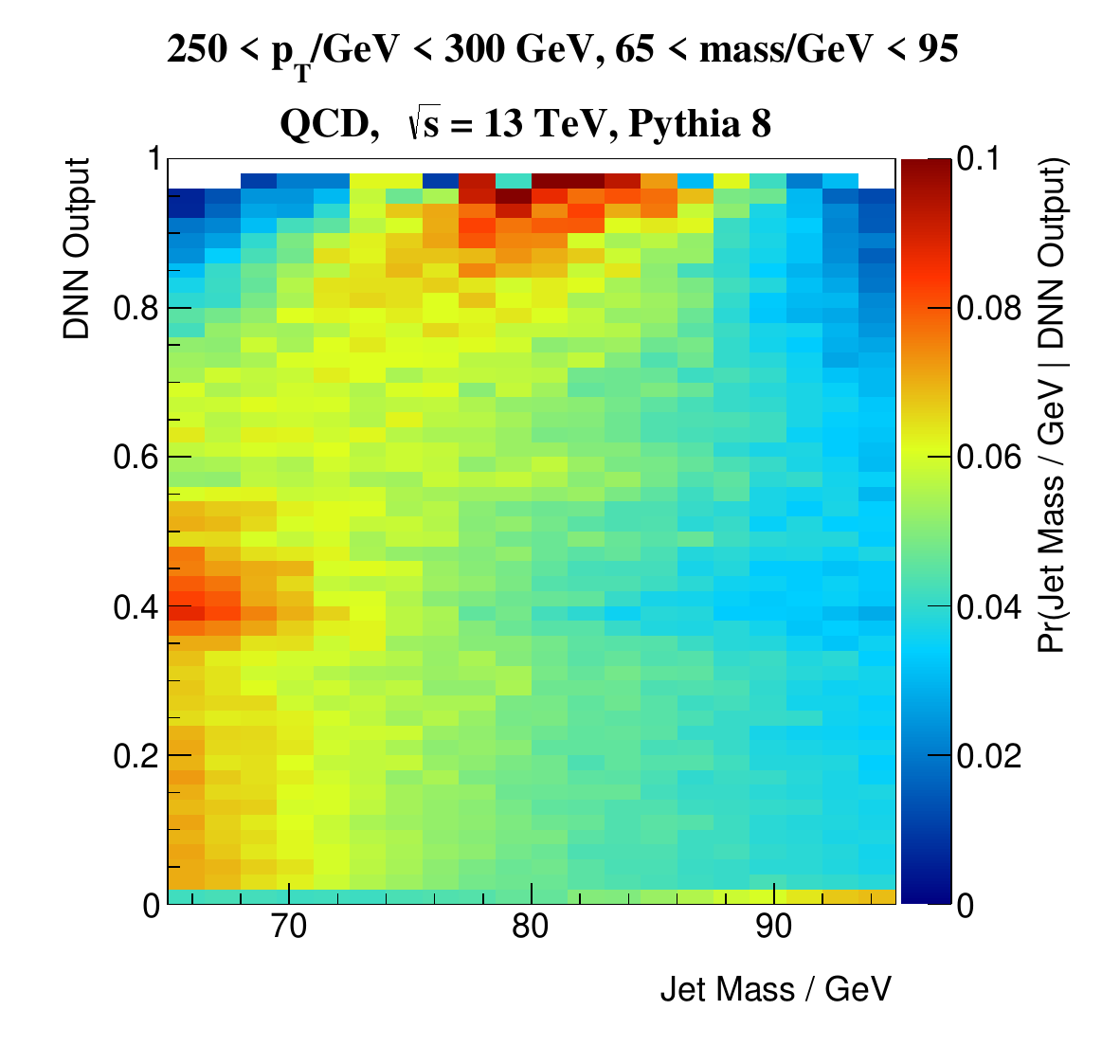}
         \includegraphics[width=0.32\textwidth]{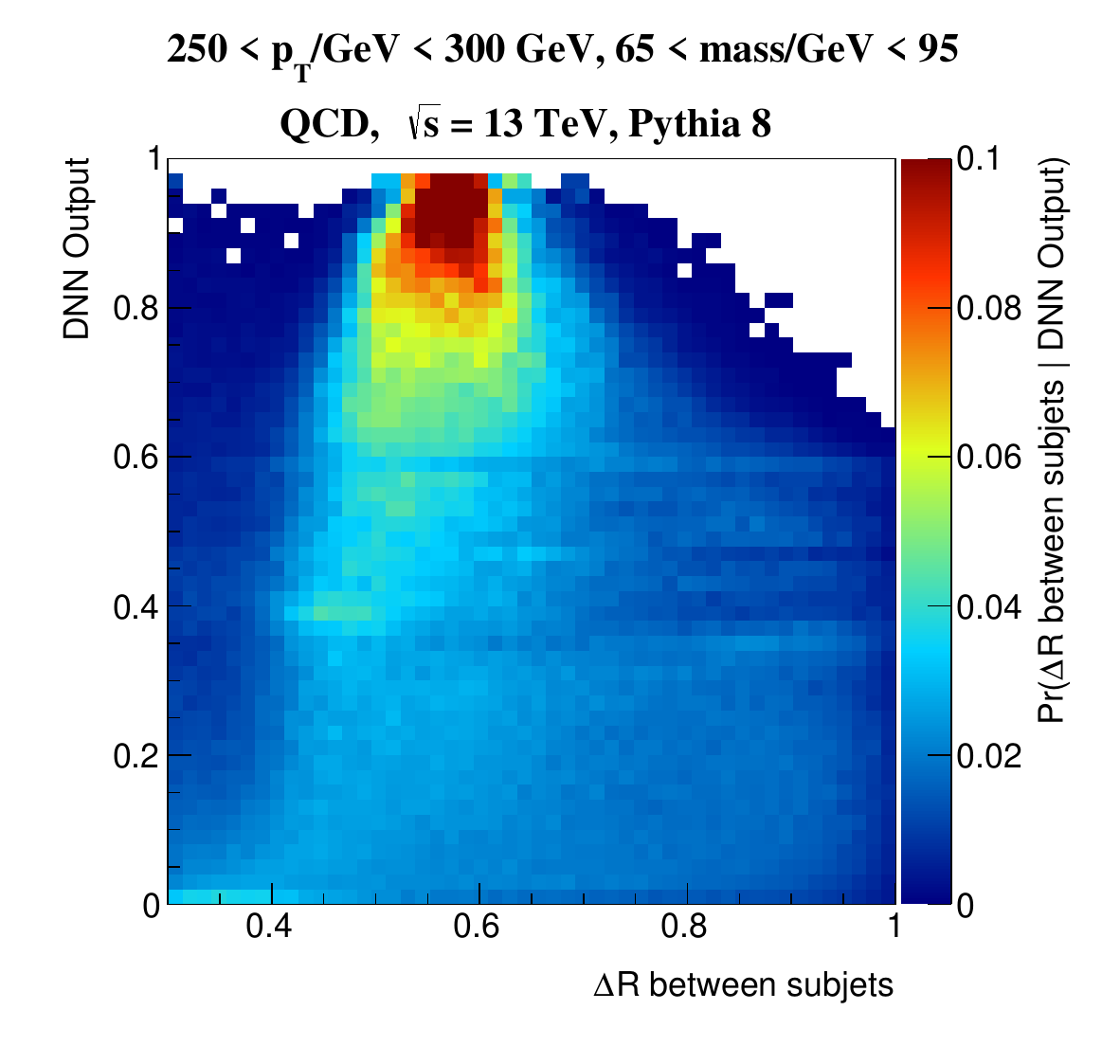} 
         \includegraphics[width=0.32\textwidth]{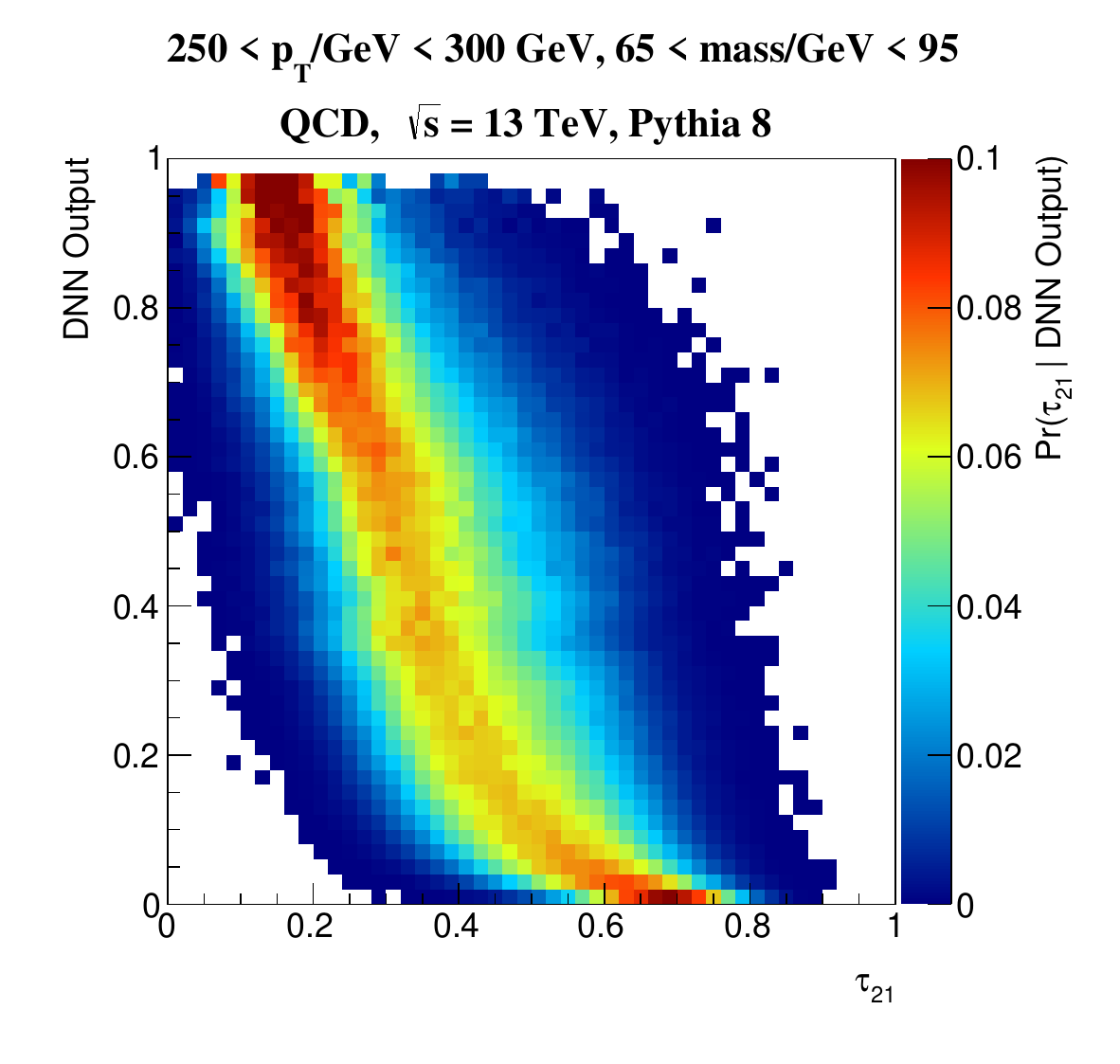}
      }\\

      \caption{Network output versus mass (left), $\Delta R$ (middle), and $\tau_{21}$ (right) for the ConvNet network (MaxOut distributions are similar).  Each row is normalized and represents the probability distribution of the variable shown on the x-axis given the network output.}
      \label{fig:qcdsculpt}

    \end{center}
\end{figure}

The conditional distributions between the DNN output and the physics-variables are shown in Figure~\ref{fig:sculptedConv} for the ConvNet network against the jet mass, $\Delta R$, and $\tau_{21}$.   These distributions are normalized in bins of the DNN output, and thus the $z$-axis shows a discretized estimate of the conditional probability density of a physics variable value given the network output (i.e. $\Pr(\text{variable}|\text{network output})$). Normalizing the distributions in this way allows us to see the most probable values of the physics variables at each point of the network output, without being affected by the overall distribution of jets in this 2D space.  There is a strong non-linear relationship between $\tau_{21}$ and $\Delta R$, giving further evidence that this information has been learned by the network.  However, the correlations are much weaker with the jet mass variable. 
While it is not shown, similar patterns are found for the MaxOut and Conv-Norm networks.   For reference, the full joint distributions can be found in Appendix~\ref{sec:app:dists}. 

\subsection{Understanding what is learned} 
\label{ssub:understanding_what_is_learned}

In order to gain a deeper understanding of the physics leaned by the DNNs, in this section we examine how the internal structure of the network relates to the substructure and properties of W bosons versus QCD jets.

\begin{figure}[bt]
  \begin{center}
      \subfloat[$(11\times11)$ convolutional kernels from first layer \label{subfig:filters}]{
        \includegraphics[width=0.45\textwidth]{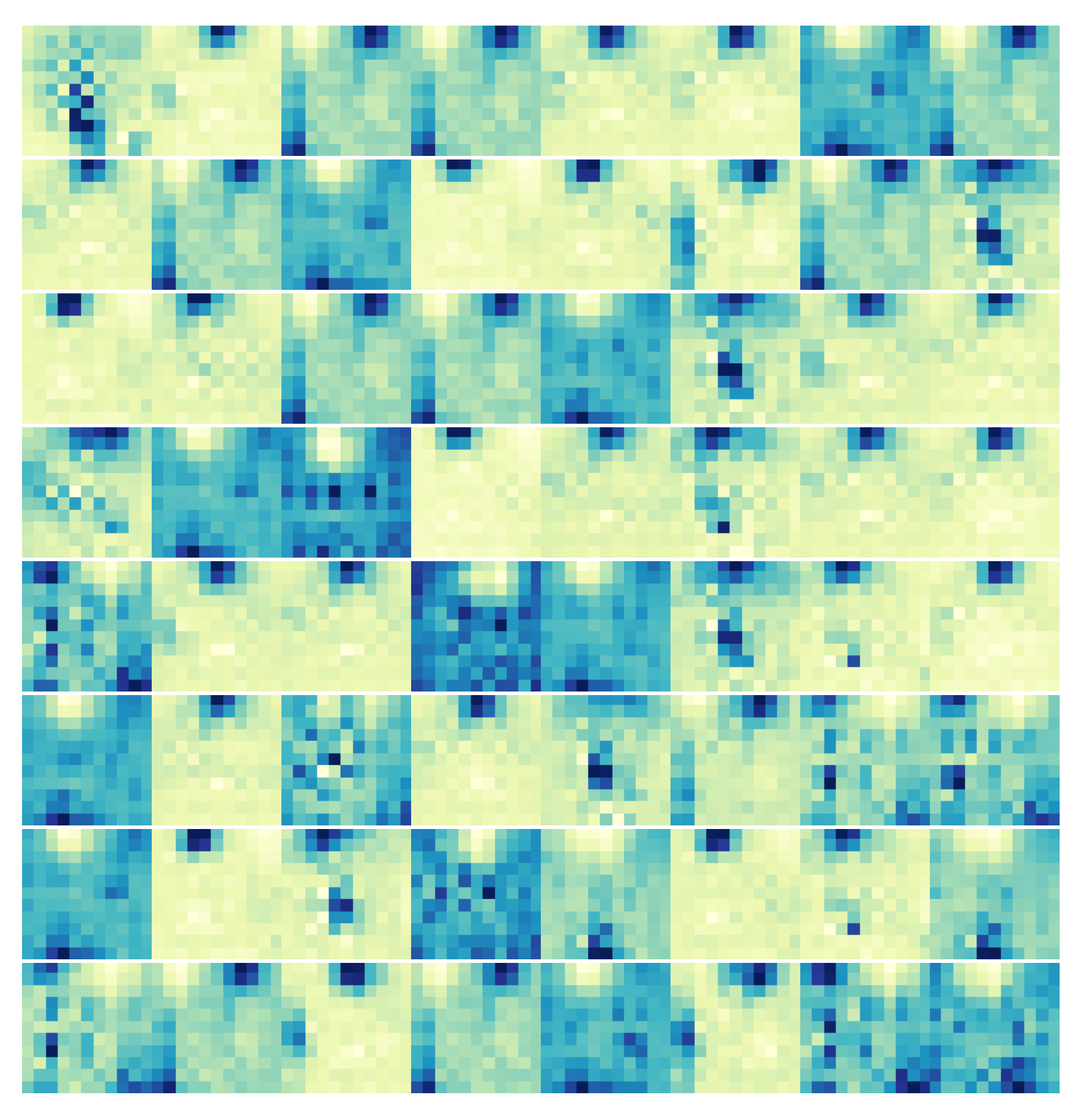}
      }
      \subfloat[Convolved Jet Image differences\label{subfig:convolvedfilters}]{
        \includegraphics[width=0.45\textwidth]{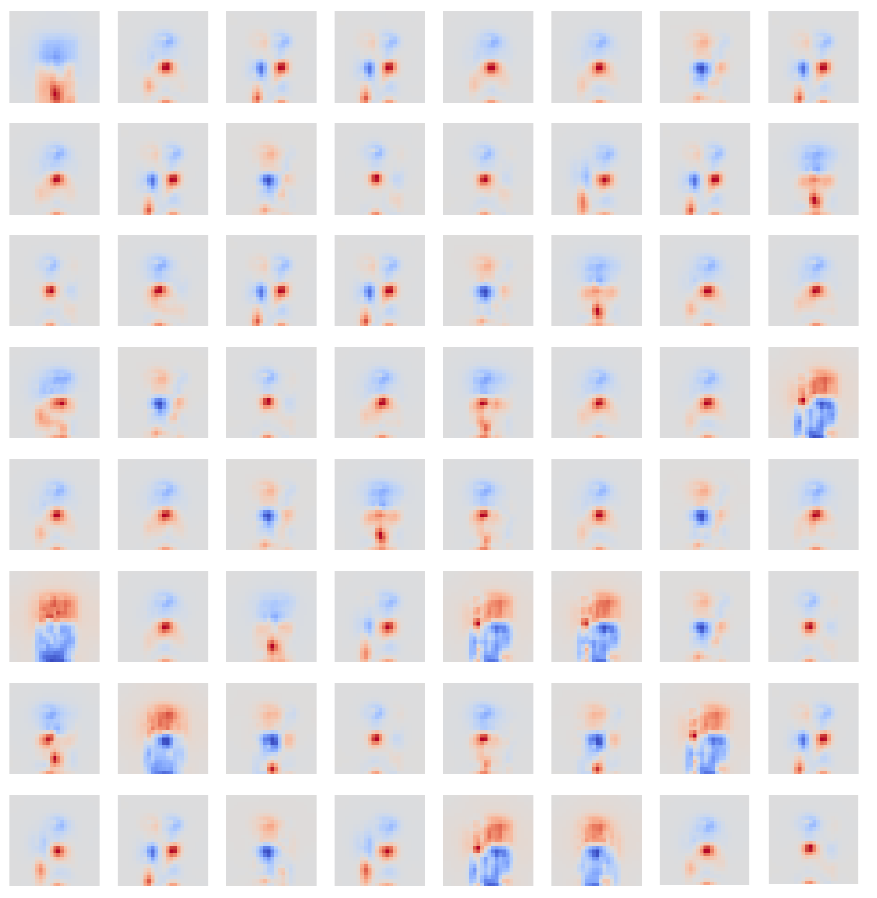}
      }
      \caption{Convolutional Kernels (left), and convolved feature differences in jet images (right)}
      \label{fig:convkernels}

    \end{center}
\end{figure}

In Figure~\ref{subfig:filters}, we show the first layer 11$\times$11 convolutional filters learned by the Conv-Norm network. Each filter is visualized by showing the learned weight in each position of the filter $W_{ij}$ from Section~\ref{sec:arch}.  We can see that there is variation between filters, indicating that they are learning different features of the jet-images, but this variation is not as large as seen in many CV problems due to the sparsity of the jet-images.  We also see that they tend to learn representations of the subjets and distances between subjets, as seen by the circular features found in many of the filters.

To get a better understanding of how these filters provide discrimination, we mimic the operation in the first layer of the network by convolving each filter with average of large samples of signal and background jet images.  The difference between the convolved average signal and background jet-images helps to provide an understanding of what difference in features the network learns at the first layer in order to help discriminate.

More formally, let $J_s=\frac{1}{n}\sum_{i:i\text{ is signal}} J^{(i)}$ and $J_b=\frac{1}{n}\sum_{i:i\text{ is background}}J^{(i)}$ represent the average signal and background jet over a sample, where $J^{(i)}$ is the $i$th jet image. In addition, we can select a filter $w_i\in\mathbb{R}^{11\times11}$ from the first convolutional layer. We then examine the differences in the post convolution layer by computing:
\begin{equation}
  J_s \ast w_i - J_b \ast w_i, \forall i,
\end{equation}

\noindent where $\ast$ is the convolution operator. We arrange these new ``convolved jet-images'' in a grid, and show in red regions where signal has a stronger representation, and in blue where background has a stronger representation. In Figure~\ref{subfig:convolvedfilters}, we show the convolved differences described above, where each $(i, j)$ image is the representation under the $(i, j)$ convolutional filter. We note the existence of interesting patterns around the regions where the leading and subleading subjets are expected to be. We also draw attention to the fact that there is a large diversity in the the convolved representations, indicating that the DNN is able to learn and pick up on multiple features that are descriptive.

\begin{figure}[!htbp]
  \centering
  \includegraphics[width=0.8\textwidth]{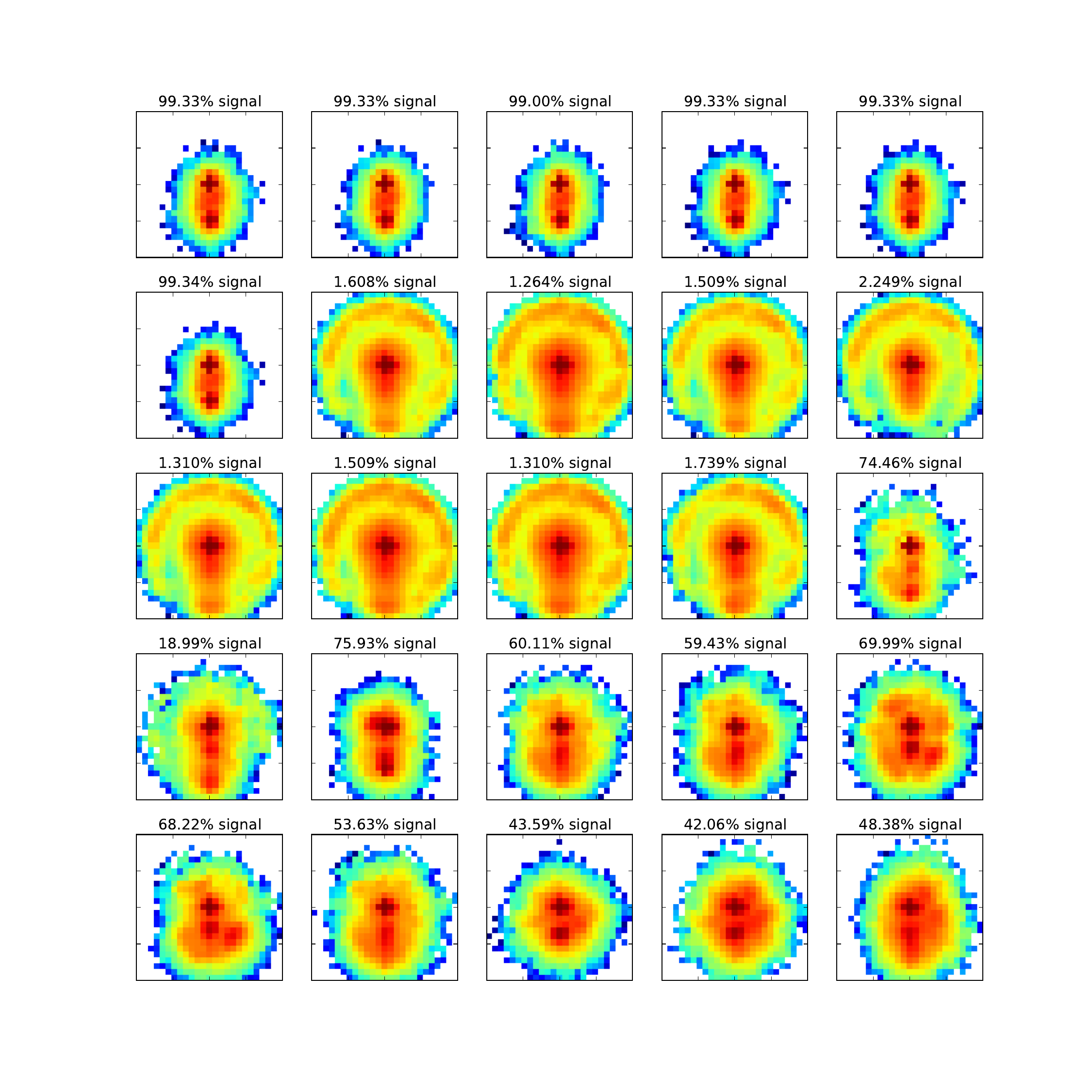}
  \caption{The average of the 500 jet images with the highest node activation  for the last hidden layer of the MaxOut network.  The nodes are ordered from top left to bottom right by increasing sparsity.  The top left is the most commonly activated node whereas the bottom right node is least activated and frequently zero. }
  \label{fig:mostactiviate}
\end{figure}

A related way to visualize the information learned by various nodes in the network is to consider the jet images which most activate a given node.  Fig.~\ref{fig:mostactiviate} shows the average of the 500 jet images with the highest node activation  for the last hidden layer of the MaxOut network (the layer before the classification layer).  The first row of images in Fig.~\ref{fig:mostactiviate} show clear two-prong signal-like structure whereas the second and third rows show one-prong diffuse radiation patterns that are more background-like.  The remaining rows have a variety of $\Delta R$ distances between subjets and have a mix of background and signal-like features.


\clearpage
\newpage

\subsection{Physics in Deep Representations} 
\label{ssub:physics_in_deep_representations}
\begin{figure}[!htbp]
  \centering
\includegraphics[width=0.5\textwidth]{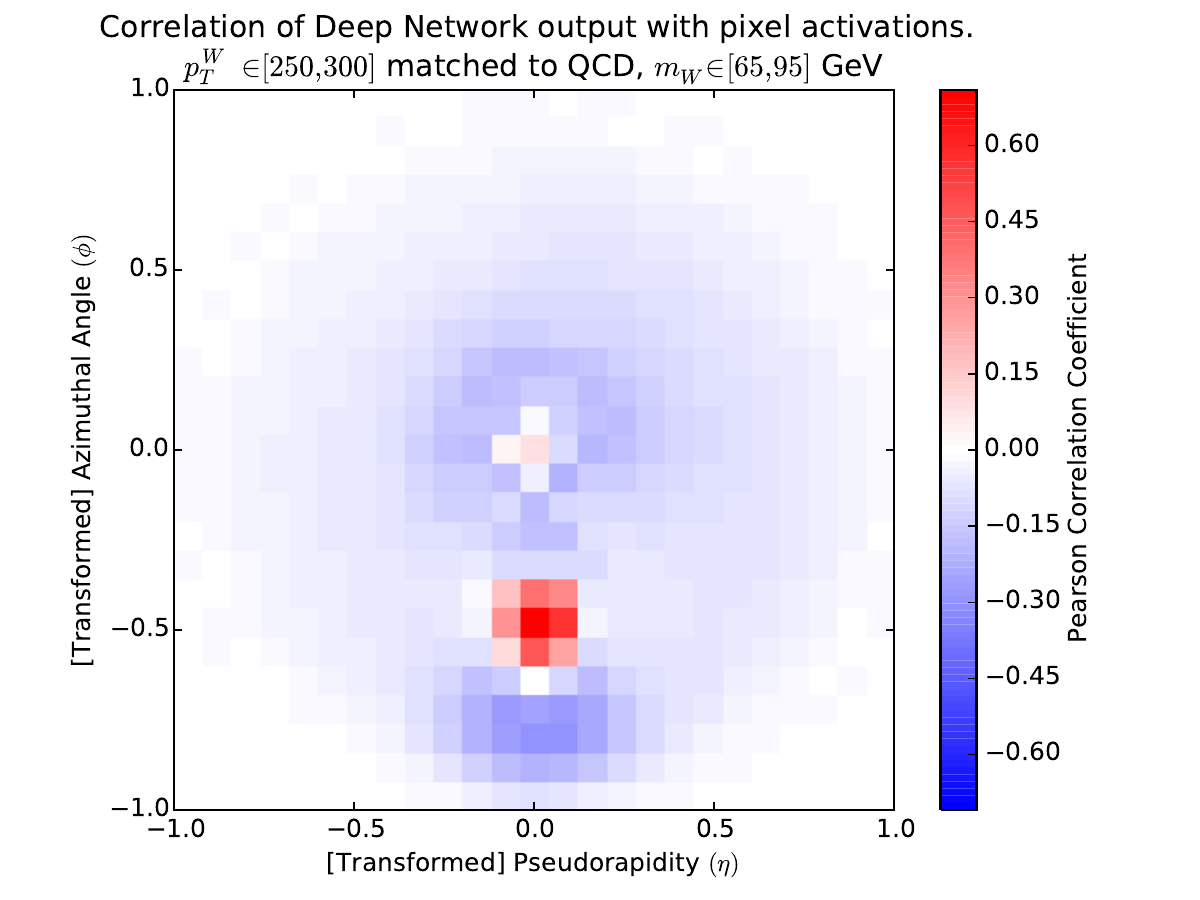}\includegraphics[width=0.5\textwidth]{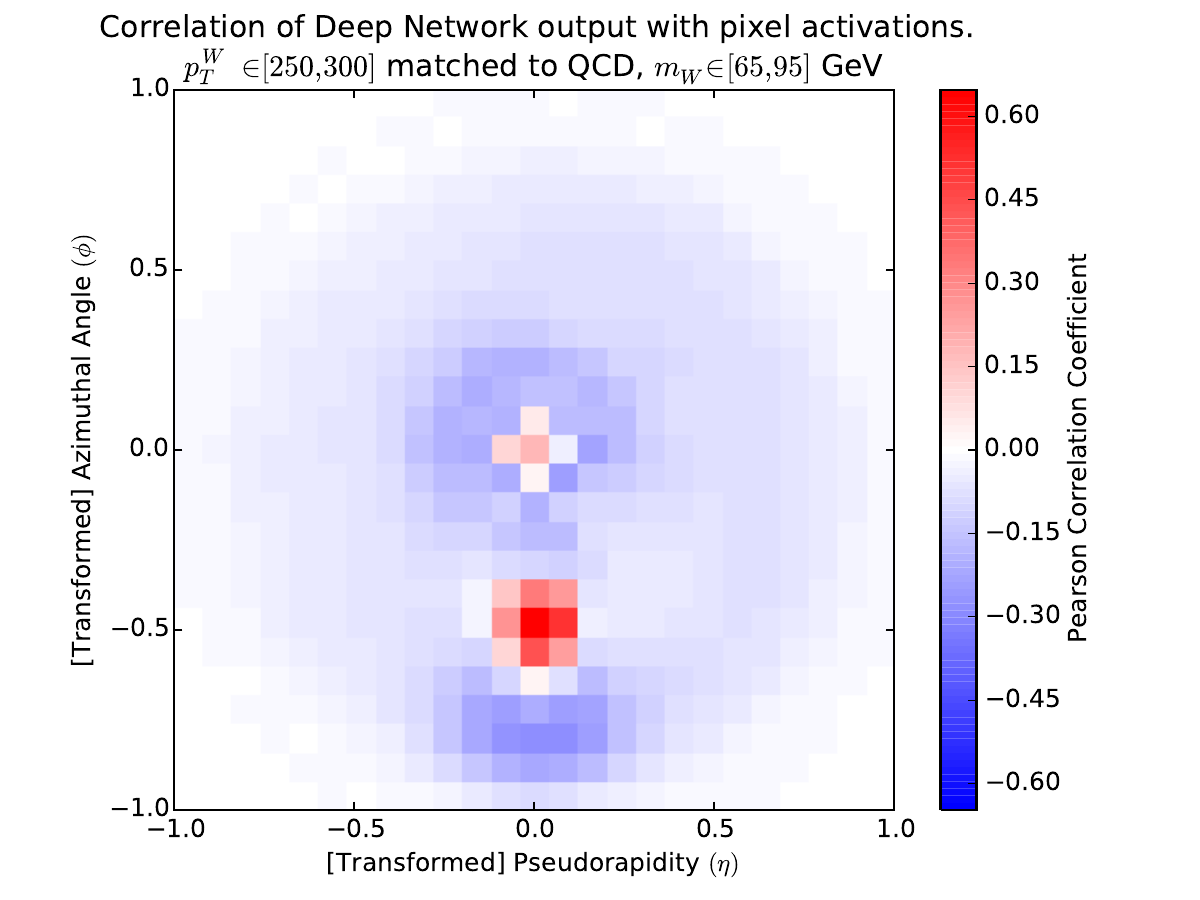}
  \caption{Per-pixel linear correlation with DNN output for the Convnet (left) and the MaxOut network (right).  Signal and background jets are combined.}
  \label{fig:corr}
\end{figure}

To get a tangible and more intuitive understanding of what jet structures a DNN learns, we compute the correlation of the DNN output with each pixel of the jet-images. Specifically, let $y$ be the DNN output, and consider the intensity of each pixel $I_{ij}$ in transformed $(\eta, \phi)$ space. We the construct an image, which we denote the \emph{deep correlation jet-image}, where each pixel $(i, j)$ is $\rho_{I_{ij}, y}$, the Pearson Correlation Coefficient of the pixels intensity with the final DNN output, across images. While this this image does not give a direct view of the discriminating information learned within the network, it does provide a guide to how such information may be contained within the network.  In Figure~\ref{fig:corr}, we construct this deep correlation jet-image for both the ConvNet and the MaxOut networks.  We can see that the location and energy of the subleading subjet, found at the bottom of the image, is highly correlated with the DNN output and important for identifying signal jet-images.  In contrast, the information contained in the leading subjet, seen at $(x,y)\sim (0,0)$ in the image, is not particularly correlated with the network output owing to the fact that both signal and background jets have high energy leading subjets.  We also see asymmetric regions around both subjets that are correlated with the DNN output and is indicating the presence of additional radiation expected in the QCD background jets.  Finally, a small negative correlation with the rest of the jet area is seen, indicating that radiation from the background jets is more likely to be observed in these regions.   The exact function form of these distribution are not known, nor does it seem to describe exactly any known physics inspired variable.

\clearpage

\subsection{Studies in the Uniform Phase Space} 
\label{sub:flat_hypercube_studies}

\begin{figure}[htbp]
  \centering
      \includegraphics[width=0.5\textwidth]{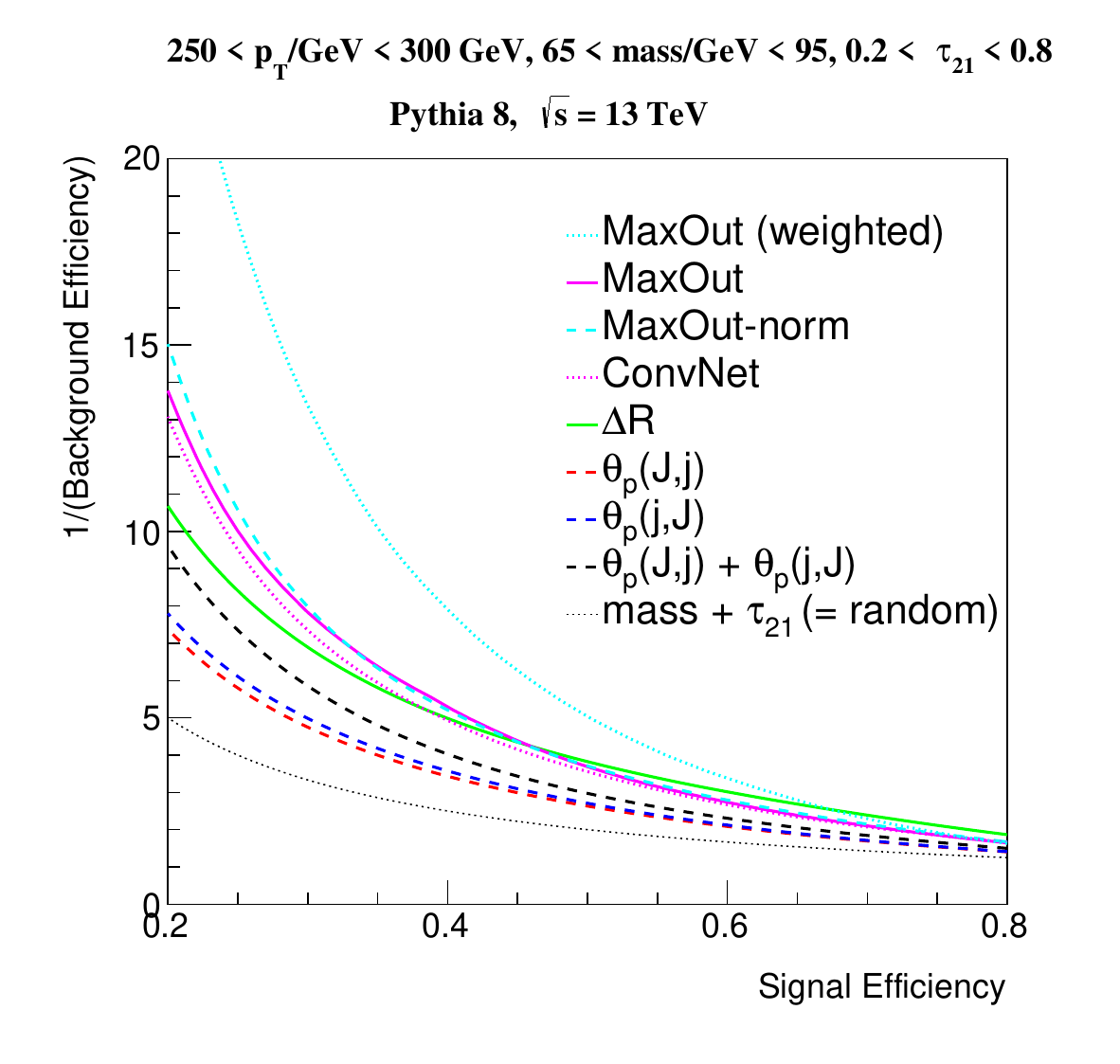}
  \caption{Various ROC curves with event weights that enforce Eq.~\ref{eq:flat} inside $m\in[65, 95]$ GeV,  $p_T\in[250, 300]$ GeV, and  $\tau_{21}\in[0.2, 0.8]$.  By construction, the $\tau_{21}$ and likelihood combination of $\tau_{21}$ and mass are non-discriminating (and are thus equal to a random guess).  The ConvNet, MaxOut, and MaxOut-Norm networks are trained without the weights applied and the MaxOut (weighted) line was trained with the weights applied during training.}
  \label{fig:rocCube}
\end{figure}

An important part of the investigation into what the neutral networks are learning beyond the standard physics features is to quantify the performance when these features are removed.  This represents the {\it unique information} learned by the network.  One way to remove the discrimination power from a given feature is to apply a transformation such that the marginal likelihood ratio is constant at unity.  In other words, we derive event-by-event weights such that

\begin{equation}
\label{eq:flat}
  f(m, \tau_{21}, p_T| W'\rightarrow WZ) \approx f(m, \tau_{21}, p_T| QCD),
\end{equation}

\noindent where $f(X|Y)$ is the probability density function of $X$ given $Y$.  This is done practically by binning the mass and $\tau_{21}$ distributions and then assigning to each event a weight given by the inverse bin content corresponding to the jet mass and $\tau_{21}$ of that particular event. Figure~\ref{fig:rocCube} shows the ROC curve for various features with this weighting scheme applied.  By construction, $\tau_{21}$ and the jet mass do not have any discrimination power between signal and background, evident by the fact that $\epsilon_\text{bkg} = \epsilon_\text{signal} = $ the random guess line.    However, the convolutional network that is trained inclusively (without the weights from Equation~\ref{eq:flat}) does have some discrimination power when the weights from Equation~\ref{eq:flat} are applied.  For a fixed signal efficiency, the overall performance is significantly degraded with respect to the un-weighted ROC curve in Figure~\ref{fig:combinedROC1}, but the improvement over a random guess is significant.  Interestingly, the network performance is significantly better in this re-weighted setting when the same weighting is applied during training (effort by the network is not needed to learn $\tau_{21}$, for instance).  The ConvNet and MaxOut procedures training inclusively have similar performance.

Figure~\ref{fig:corr} already suggested that information about {\it colorflow} is contributing to the performance of the tagger since the signal is a color singlet and the background is predominantly a color octet (gluon).  The radiation pattern in the former case is expected to be concentrated between the subjets of the jet and in the latter case around the subjets.  One variable designed~\cite{Gallicchio:2010sw} and recently shown~\cite{Aad:2015lxa} to be sensitive to the colorflow is the jet pull angle, $\theta_P(j_1,j_2)$ for jets $j_1$ and $j_2$.  The jet pull vector is given by $\vec{v}_p^{j}=\frac{1}{p_T^j}\sum_{i\in j} p_T^i|\vec{r}_i|\vec{r}_i$, where $i$ runs over the jet's constituents and $r_i$ is the vector in $(y,\phi)$ that points from the jet axis to the constituent $i$.  The pull angle  $\theta_P(j_1,j_2)$ is the angle the pull vector of jet $j_1$ makes with respect to the vector in $(y,\phi)$ pointing from the $j_1$ jet axis to the $j_2$ jet axis.  Note that $\theta_P(j_1,j_2)\neq \theta_P(j_2,j_1)$ because the former uses the substructure of $j_1$ and the latter uses the substructure of $j_2$.  We adapt the pull angle to the case of large-radius trimmed jets by using the leading ($J$) and subleading ($j$) subjets.  The red and blue dashed lines in Fig.~\ref{fig:rocCube} show that a significant fraction of the DNNs performance can be explained by colorflow information contained within the jet pull angles.  However, especially for the network trained with the weights, the DNN performance is also significantly better than the jet pull angles.

\begin{figure}[htbp]
  \centering
  \includegraphics[width=0.45\textwidth]{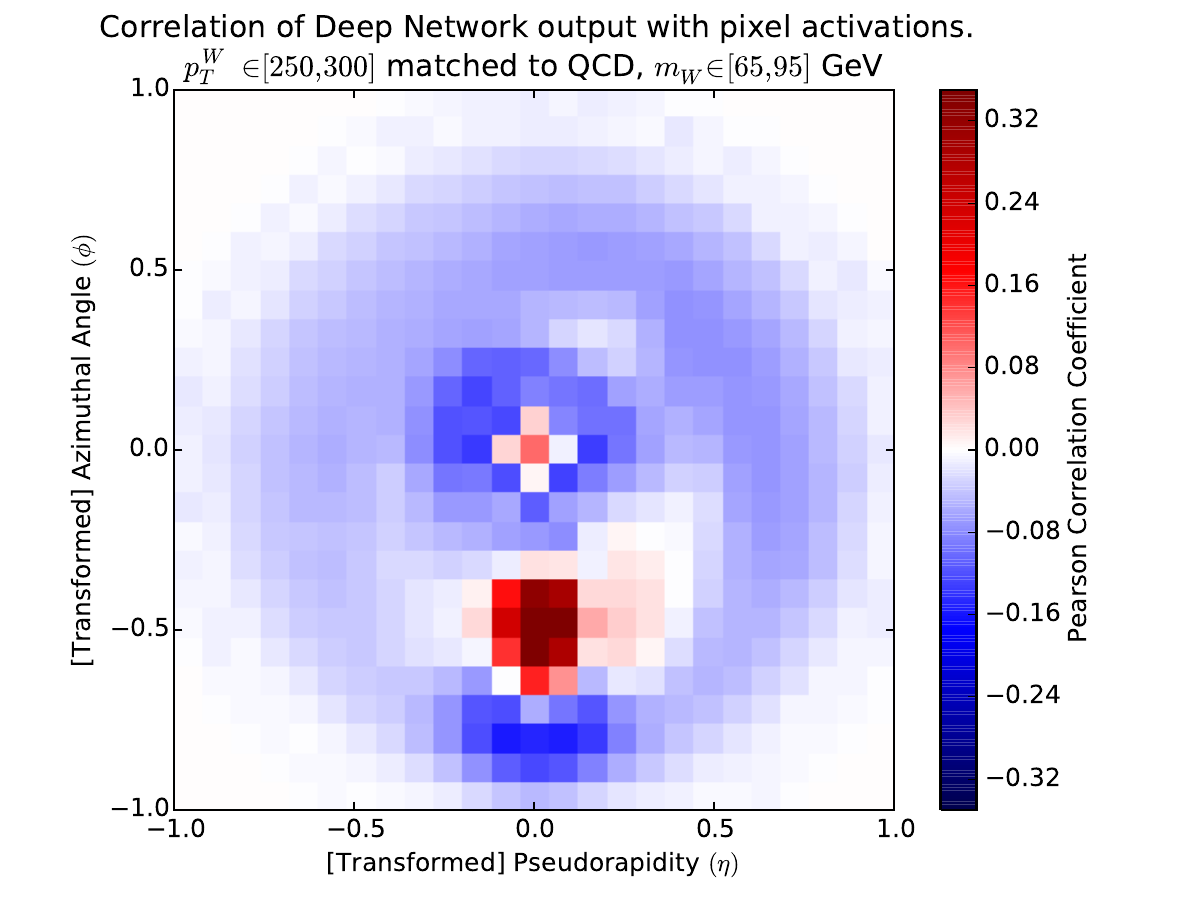} \includegraphics[width=0.45\textwidth]{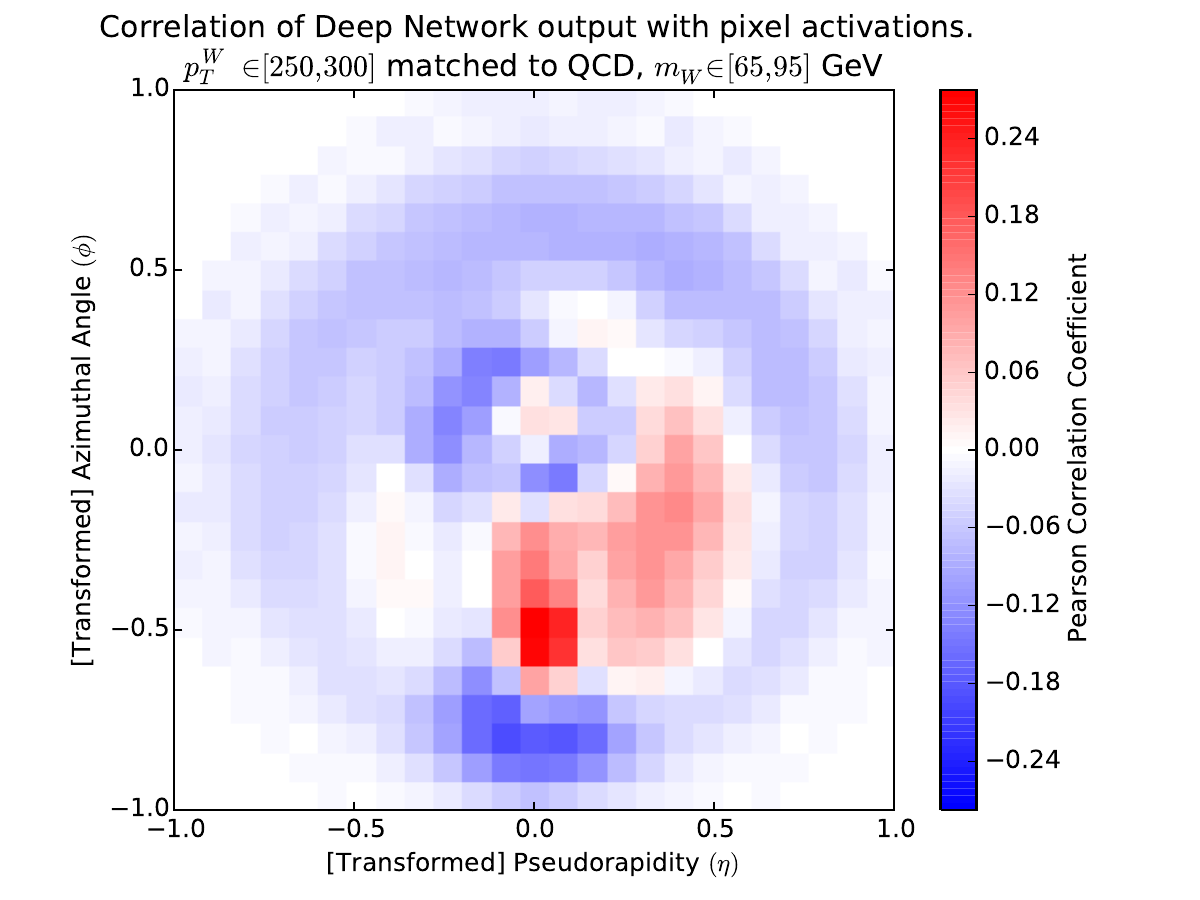}
  \caption{Pearson Correlation Coefficient for pixel intensity and the convolutional neural network output for $W'\rightarrow WZ$ and QCD (combined) for the MaxOut network training inclusively and then weighted (left) and for the MaxOut network training with the weights from Equation ~\ref{eq:flat} applied also during the training.}
  \label{fig:cor_hyper}
\end{figure}

One can gain intuition about the unique information learned by the network by studying the correlation of the network output and the pixel intensities with the Equation~\ref{eq:flat} weights applied.  This is shown in Figure~\ref{fig:cor_hyper} with and without the weights applied during training.  The two correlation plots are qualitatively similar, but the region to the right of the subjets is more enhanced when the weights are applied during the training.  This suggests that information about radiation surrounding the subjets contains important discrimination power contributing to the network's unique information.

\clearpage

\clearpage
\newpage

\subsection{Studies in the Highly Restricted Phase Space} 
\label{sub:small_window_studies}

\begin{figure}[h!]
  \begin{center}
  
        \includegraphics[width=0.99\textwidth]{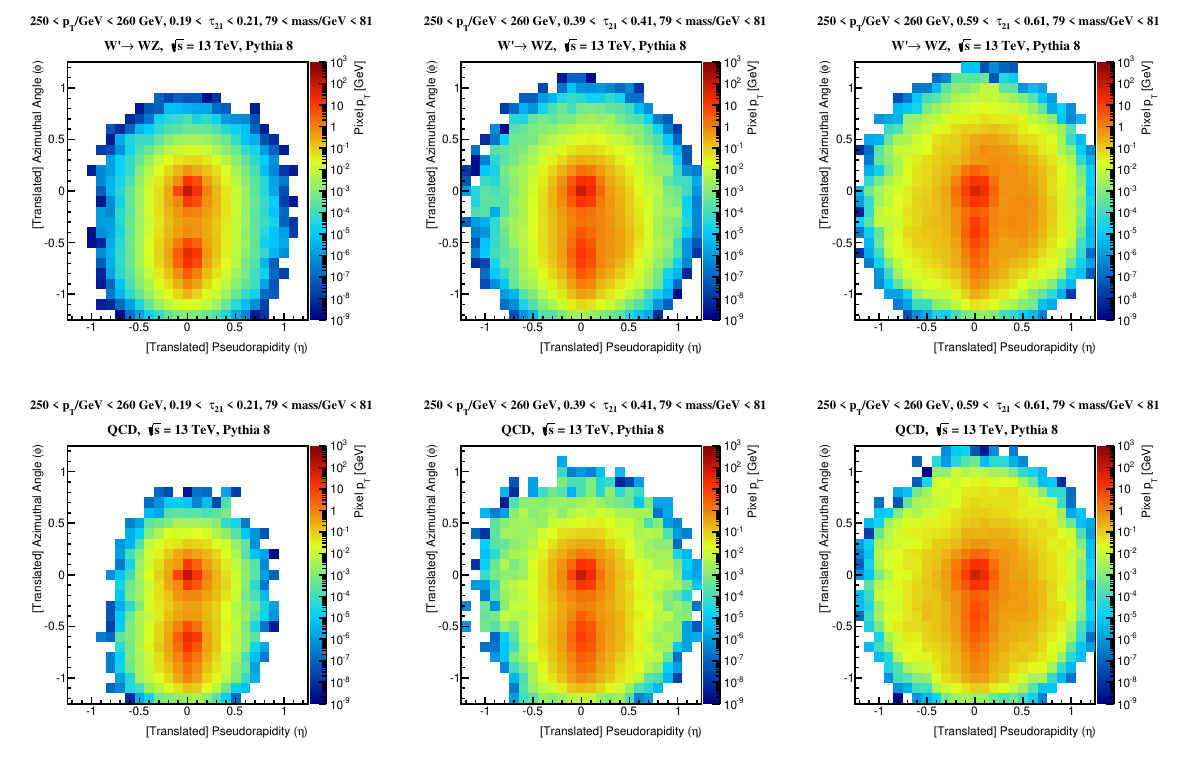}

      \caption{
        $W'\rightarrow WZ$ (top) and QCD (bottom) average jet-images in three small windows of $\tau_{21}$: [0.19, 0.21] (left), [0.39, 0.41] (middle), and [0.59, 0.61] (right).  In all cases, jet mass is restricted to be between 79 GeV and 81 GeV and the jet $p_T$ is required to be in the interval [250,260] GeV.
        \label{fig:meanImagesWindow} 
      }
    \end{center}
\end{figure}

 Another way to quantify the unique information learned by the network that also provides useful information about physical information learned by the network is to restrict the considered phase space such that $\tau_{21}$ and the jet mass distributions do not vary appreciably over the reduced space.  Figure~\ref{fig:meanImagesWindow} shows the average signal and background jet image in three small windows of $\tau_{21}$, jet mass, and jet $p_T$.  In all three windows, the jet mass is restricted to be between 79 GeV and 81 GeV and the jet $p_T$ is required to be in the interval [250,260] GeV.  The three windows are then defined by their value of $\tau_{21}$: [0.19,0.21] in the most two-prong-like case, [0.39,0.41] in a region with likelihood ratio near unity and [0.59,0.61] in a mostly one-prong-like case.  The key physics features of the jets falling in these windows are easily visualized from the average jet images.  The most striking observation is that in these three windows, signal jets look very similar to background jets.  When $\tau_{21}\in[0.19,0.21]$, both signal and background jets have a second subjet that is distinct from the leading subjet, which becomes less prominent as the value of $\tau_{21}$ increases.  

\begin{figure}[h!]
  \begin{center}
  
        \includegraphics[width=0.99\textwidth]{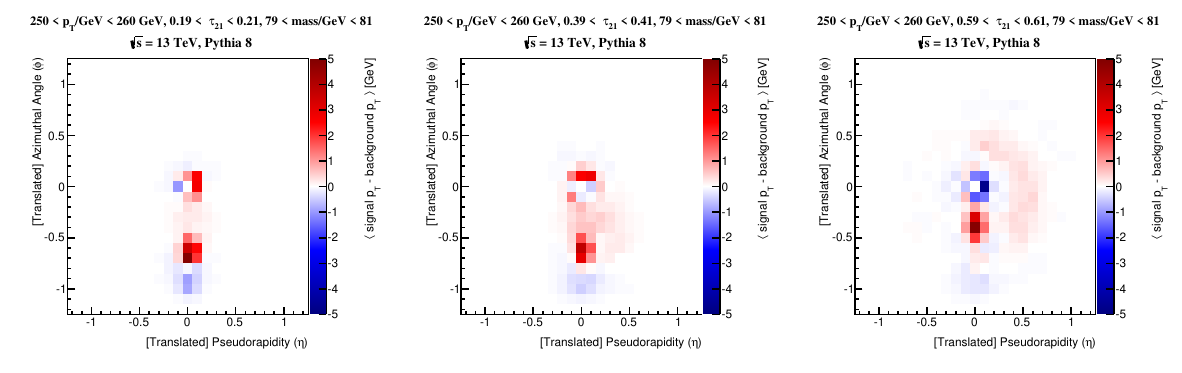}
        
      \caption{
         The average difference between $W'\rightarrow WZ$ jet-images in three small windows of $\tau_{21}$: [0.19, 0.21] (left), [0.39, 0.41] (middle), and [0.59, 0.61] (right).  In all cases, jet mass is restricted to be between 79 GeV and 81 GeV and the jet $p_T$ is required to be in the interval [250,260] GeV.  The red colors are more signal-like and the blue is more background-like.
        \label{fig:meanImagesWindow2} 
      }
    \end{center}
\end{figure}

The differences between images in these small windows tells us about what information {\it could be learned} by the networks beyond $\tau_{21}$ and the jet mass.  Since the differences are subtle, the average difference is explicitly computed and plotted in Figure~\ref{fig:meanImagesWindow2} for the three narrow windows of $\tau_{21}$.  In the window with $\tau_{21}\in$[0.19,0.21], there are five features: a localized blue patch in the bottom center, a localized red patch just above that, a red diffuse region between the red patch and the center and then a blue dot just left of center surrounded by a red shell to the right.  Each of these have a physics meaning: the lower two localized patches give information about the orientation of the second subjet ($\Delta R$) which is slightly wider for the QCD jets which need a slightly wider angle to satisfy the mass requirement.  The red diffuse region just above the localized patches is likely an indication of colorflow as introduced earlier: the $W$ bosons are color singlets compared to the color octet gluon jet background, and thus we expect the radiation pattern to be mostly between the two subjets for the $W$.  One can draw similar conclusions for all the features in each of the plots in Figure~\ref{fig:meanImagesWindow2}.

\begin{figure}[h!]
  \centering
   \includegraphics[width=0.65\textwidth]{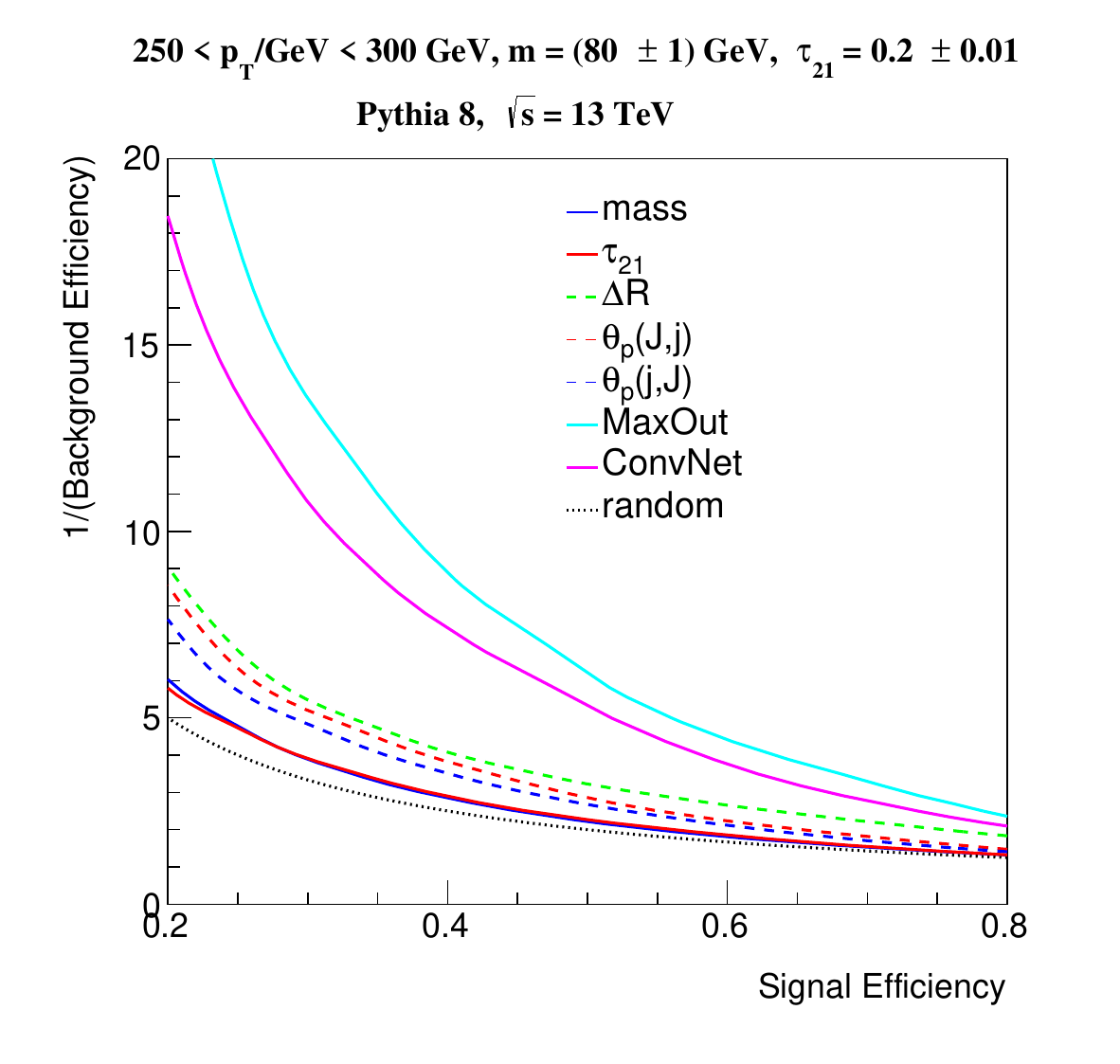}
  \caption{ROC curves for $m_\text{jet}\in [79, 81]$ GeV, $p_{T}\in [250, 255]$ GeV, $\tau_{21}\in[0.19, 0.21]$. By construction, $\tau_{21}$ is no better than a random guess in this small window.  The neural networks are trained inclusively (but still within the stated mass and $p_T$ windows).}
  \label{fig:rocWindow}
\end{figure}


Now, we turn back to the neutral network and their performance in these small windows of jet mass and $\tau_{21}$.    Figure~\ref{fig:rocWindow} shows three ROC curves in the window $\tau_{21} \in$[0.19,0.21].  By construction, the $\tau_{21}$ and jet mass curves are not much better than a random guess, since these variables do not significantly vary over the small window.    The other curves show the performance of $\Delta R$ and the ConvNet and MaxOut neural networks trained inclusively, which have similar performance to each other.   As in the previous section, this allows us to quantify the unique information in the neural network.   Figure~\ref{fig:rocWindow} also includes the jet pull angle introduced in the context of Fig.~\ref{fig:rocCube}.  As with the earlier figure, the jet pull angles do provide useful discriminating information in this small region of phase space, but cannot account for the entire performance from the DNNs.

\begin{figure}[h!]
  \centering
  \includegraphics[width=0.3\textwidth]{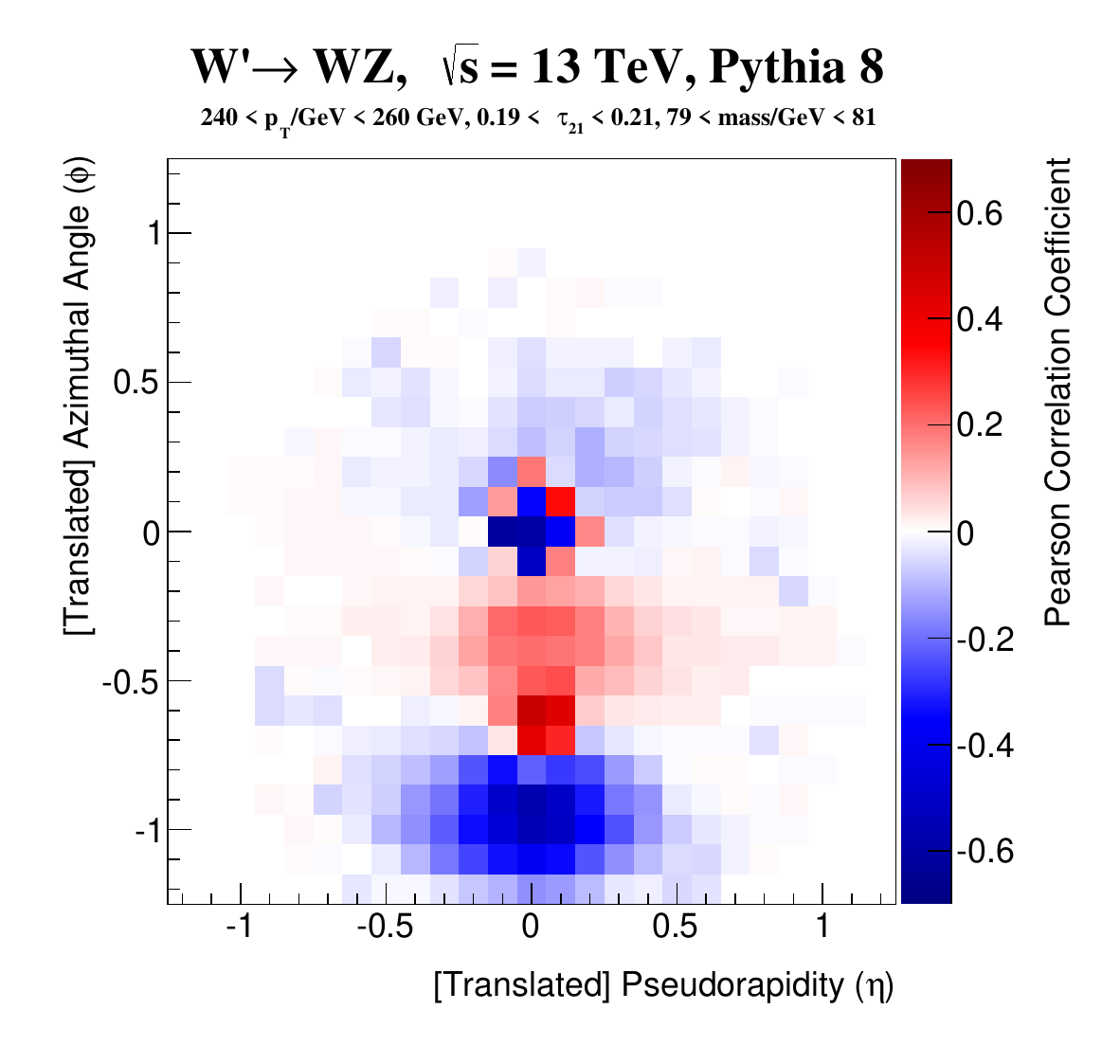}  \includegraphics[width=0.3\textwidth]{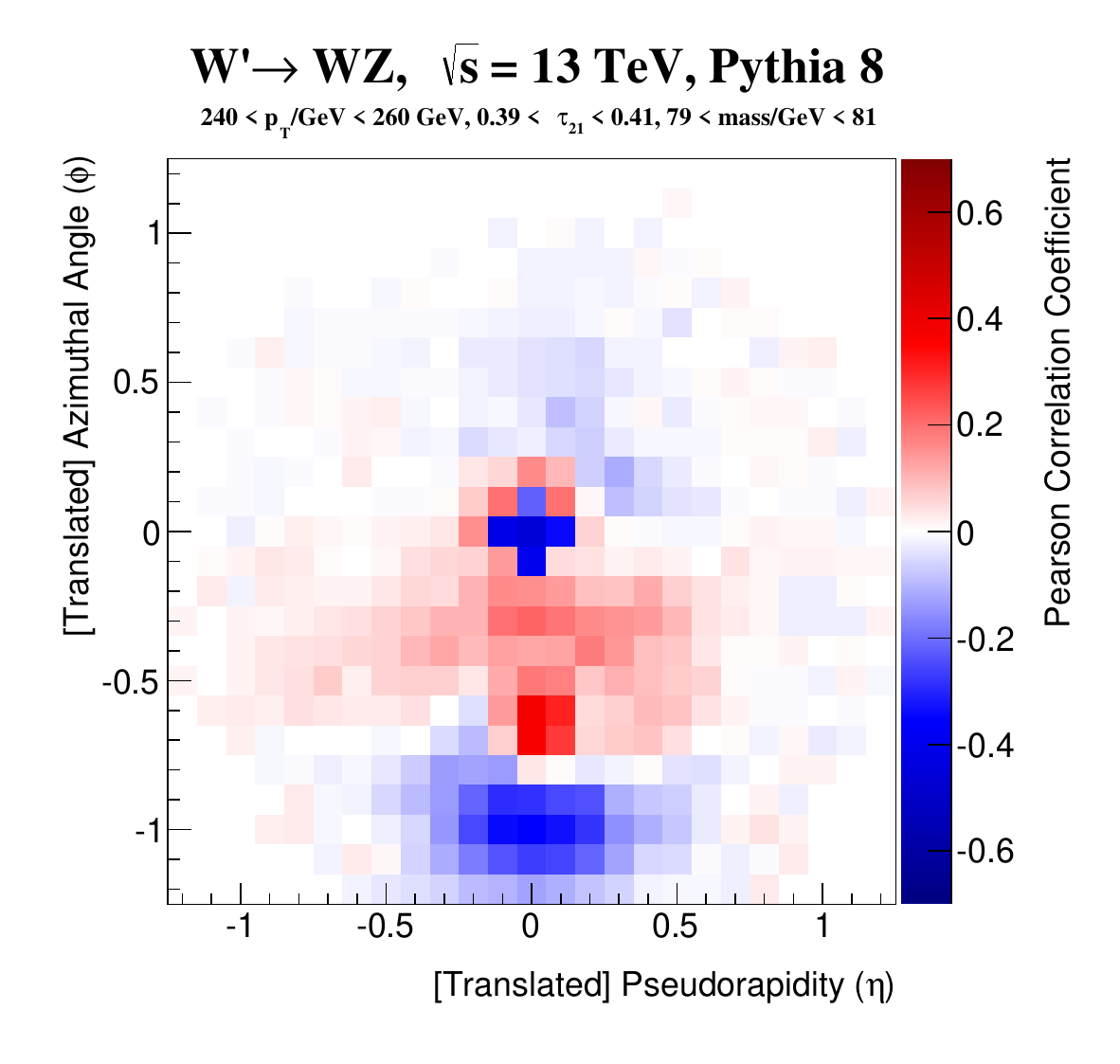}  \includegraphics[width=0.3\textwidth]{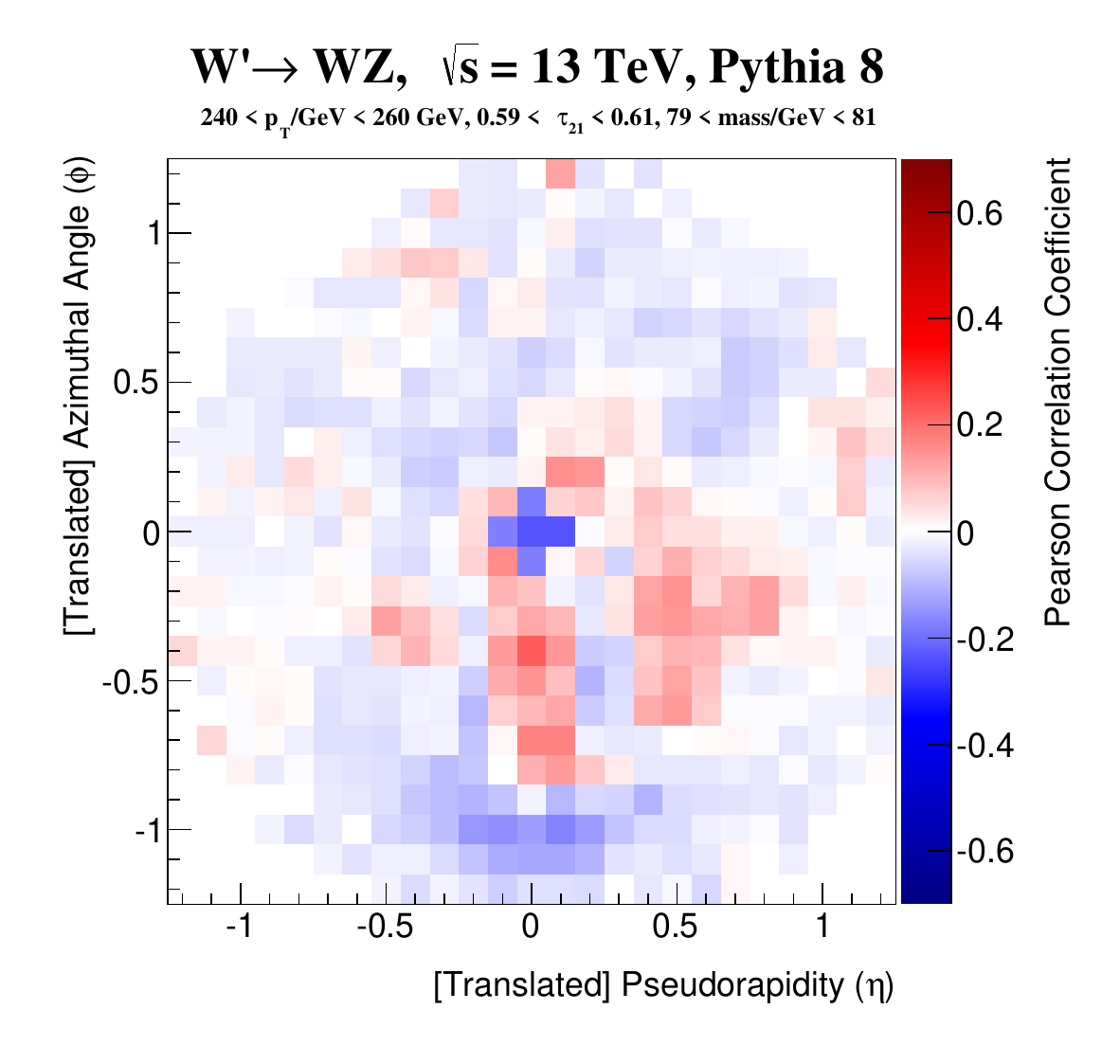}
  \caption{Pearson Correlation Coefficient for pixel intensity and the convolutional neural network output for $W'\rightarrow WZ$ and QCD (combined) in three small windows of $\tau_{21}$: [0.19, 0.21] (left), [0.39, 0.41] (middle), and [0.59, 0.61] (right).  In all cases, jet mass is restricted to be between 79 GeV and 81 GeV and the jet $p_T$ is required to be in the interval [250,260] GeV.}
  \label{fig:corrWindow}
\end{figure}

One way to {\it visualize} the unique information is to look at the per-pixel correlation between the intensity and neural network output (Figure~\ref{fig:corrWindow}).  The physical interpretation of the red and blue areas in Figure~\ref{fig:corrWindow} are related to the colorflow of $W$ and background jets.  The area in-between the subjets should have more radiation than the area around and outside of the subjets for $W$ jets and vice-versa for QCD jets.  While Figure~\ref{fig:corrWindow} is not directly the discriminant used in the network and only represents linear correlations with the network output, it does show non-linear spatial information and gives a sense of where in the image the network is looking for discriminating features.  Some of this information is contained in the jet pull angles, but the DNN must be learning additional information (Fig.~\ref{fig:rocWindow}).



%



\clearpage
\newpage


\section{Outlook and Conclusions}
\label{sec:conclusion}
Jet Images are a powerful paradigm for visualizing and classifying jets.  We have shown that when applied directly to jet images, deep neural networks are a powerful tool for identifying boosted hadronically decaying $W$ bosons from QCD multijet processes.  These advanced Computer Vision algorithms outperform several known and highly discriminating engineered physics-inspired features such as the jet mass and $n$-subjettiness, $\tau_{21}$.  Through a variety of studies, we have shown that some of these features are {\it learned} by the network.  However, despite detailed studies to preserve the jet mass, this important variable seems to not be fully captured by the neural networks studied in this article.  Understanding how to fully learn the jet mass is a goal of our future work.

In this paper, we propose several techniques for quantifying and visualizing the information learned by the DNNs, and connect these visualizations with physics properties.  This is studied by removing the information from jet mass and $\tau_{21}$ through a re-weighting or redaction of the phase space.  In this way, we can evaluate the performance of the network beyond these features to quantify the unique information learned by the network.  In addition to quantifying the amount of additional discrimination achieved by the network, we also show how the new information can be visualized through through the deep correlation jet image which displays the network output correlation with each input pixel.  These visualizations are a powerful tool for understanding what the network is learning.  In this case, colorflow patterns suggest that at least part of the unique information comes from the octet versus singlet nature of $W$ bosons and gluon jets. However, not all of the information is contained in well-known physically motivated color-flow-sensitive features like the jet pull angle.  The visualizations may even be useful in the future for engineering other simple variables which may be able to match the performance of the neural network.  

Both ATLAS and CMS have collected and will continue to collect large datasets filled with SM sources of boosted top quarks and $W$ bosons.  The collaborations have shown that event selections targeting these objects can be used to determine the systematic uncertainties of both simple and complex jet tagging techniques~\cite{Aad:2015rpa,Aad:2016pux,Khachatryan:2014vla,CMS:2014fya}.  These techniques can be readily adapted for the jet images DNN tagger as a first step toward applying the tools developed in this paper to improve tagging performance in practice.  Additionally, both ATLAS and CMS have achieved a better spatial resolution than their $0.1\times 0.1$ hadronic calorimeter granularity.  Figures~\ref{fig:preprocess3} and~\ref{fig:combinedROC1} show that the DNN tagger presented in this paper significantly out-performs the unpixelated jet mass.  The DNN tagger would do no worse than its stated performance with $0.1\times 0.1$ granularity because one can always down-sample the images before processing.  With more information available to the network, it is likely the DNN tagger could do even better.  Taking into account the non-uniform detector granularity in order to reduce the feature size is therefore an interesting direction of future work in adapting the methods presented here to a particular detector.

This edition of the study of jet images has built a new link between particle physics and computer vision by using state of the art deep neural networks for classifying high-dimensional high energy physics data.  By processing the raw jet image pixels with these advanced techniques, we have shown that there is a great potential for jet classification.  Many analyses at the LHC use boosted hadronically decaying bosons as probes of physics beyond the Standard Model and the methods presented in this paper have important implications for improving the sensitivity of these analyses.  In addition to improving tagging capabilities, further studies with deep neural networks will help us discover new features to improve our understanding and improve upon existing features to fully capture the wealth of information inside jets.

\section{Acknowledgements} 
\label{sec:acknowledgements}

We would like to thank Andrew Larkoski for useful conversations about the physics observed in the jet images.  This work is supported by the Stanford Data Science Initiative and by the US Department of Energy (DOE) under grant DE-AC02-76SF00515. BN is supported by the NSF Graduate Research Fellowship under Grant No. DGE-4747 and by the Stanford Graduate Fellowship.  


\appendix

\clearpage
\newpage

\section{Image Sparsity}
\label{sec:sparsity}

\begin{figure}[htbp!]
  \begin{center}
        \includegraphics[width=0.5\textwidth]{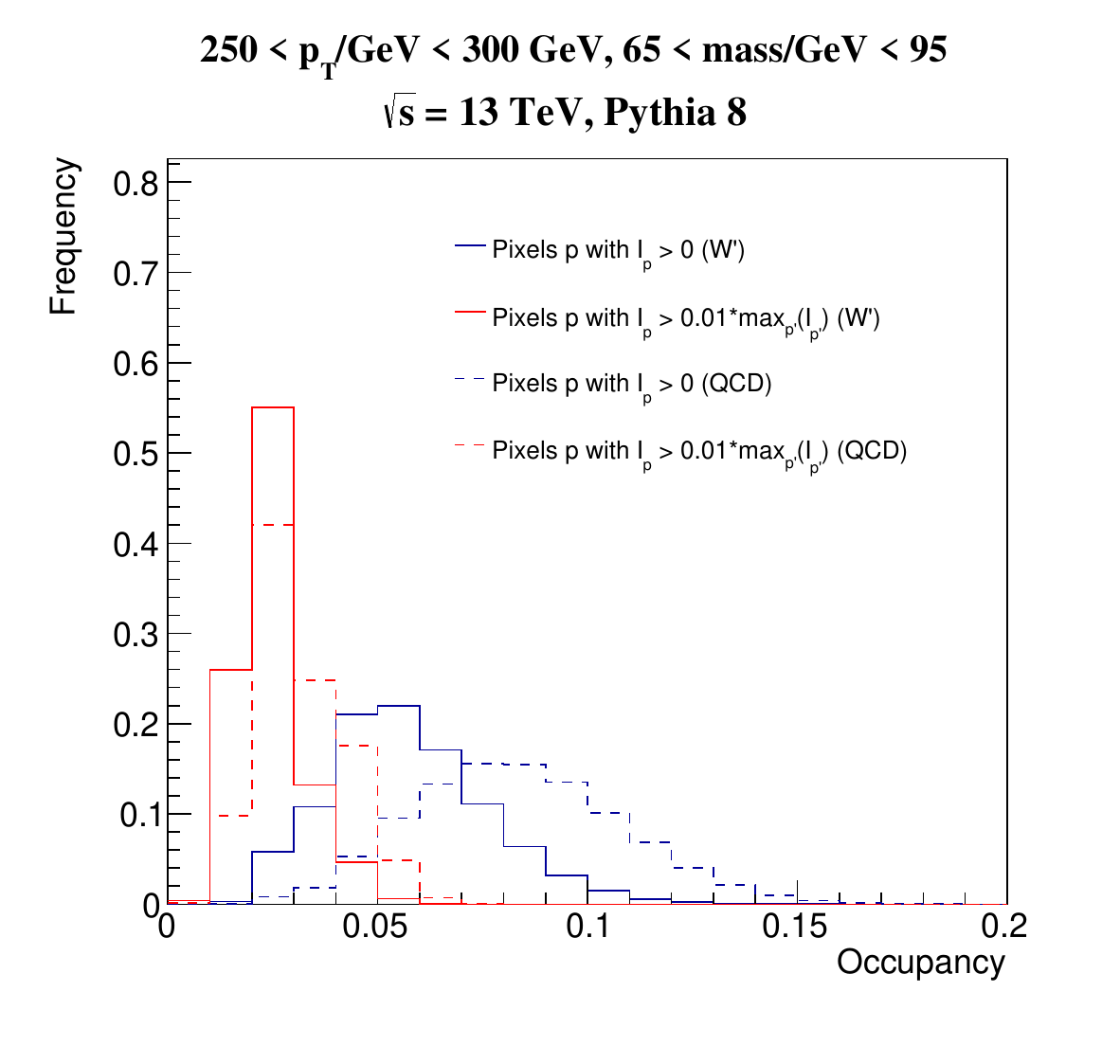}
      \caption{ The distribution of the fraction of pixels (occupancy) that have a nonzero entry (blue) or at least 1\% of the scalar sum of the pixel intensities from all pixels (red).
      \label{fig:occupancy} }
    \end{center}
\end{figure}

Figure~\ref{fig:occupancy} quantifies the sparsity of the jet images by showing the distribution of the pixel occupancy: the fraction of pixels that have a non-zero entry.  Also plotted is the fraction of pixels that have at least 1\% of the intensity of the scalar sum of the pixel intensities from all pixels.  In general, the background has a more diffuse radiation pattern and thus the corresponding jet images have a higher average occupancy.

\clearpage
\newpage

\section{Joint and Marginal Distributions}
\label{sec:app:dists}

\begin{figure}[htbp!]
  \begin{center}
        \includegraphics[width=0.45\textwidth]{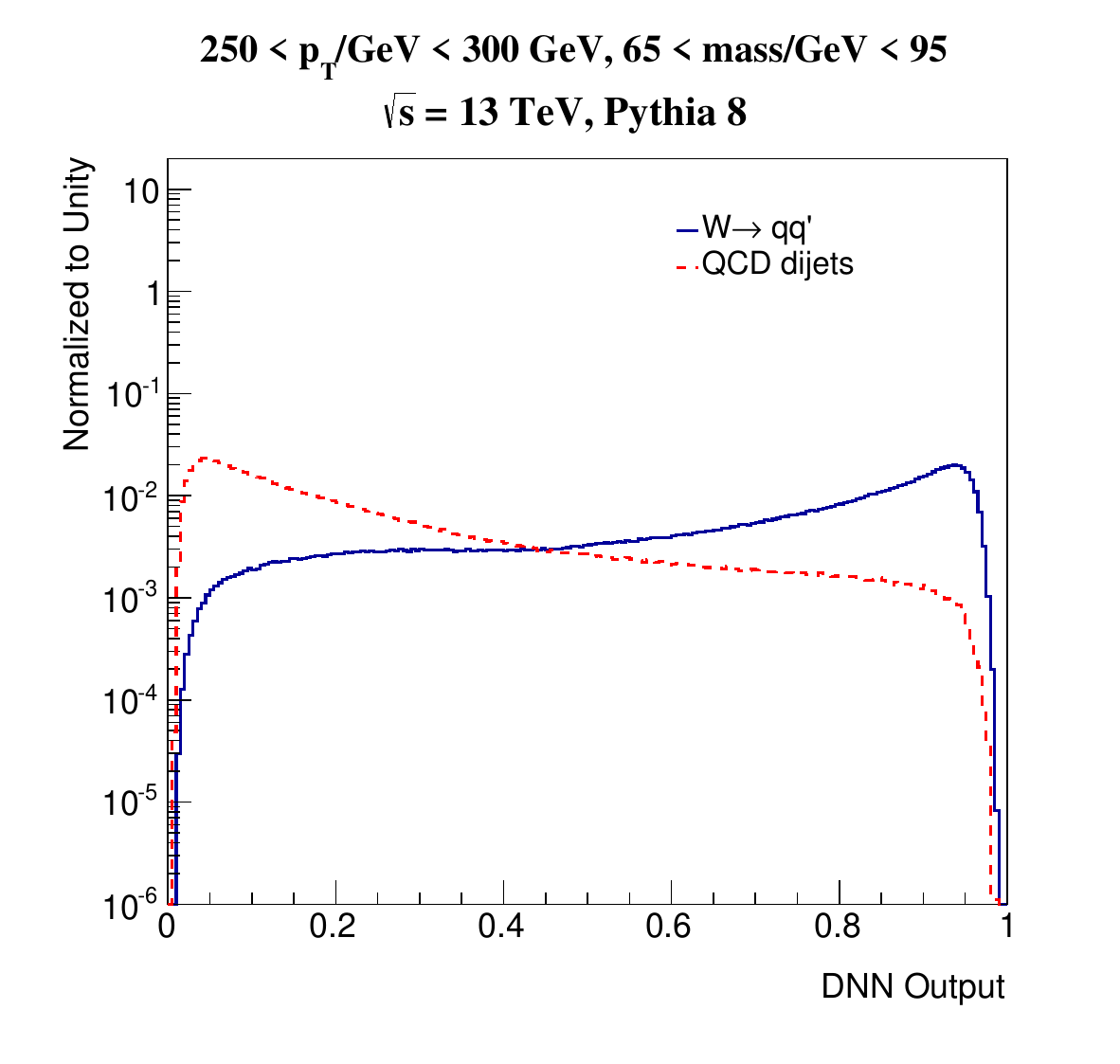} \includegraphics[width=0.45\textwidth]{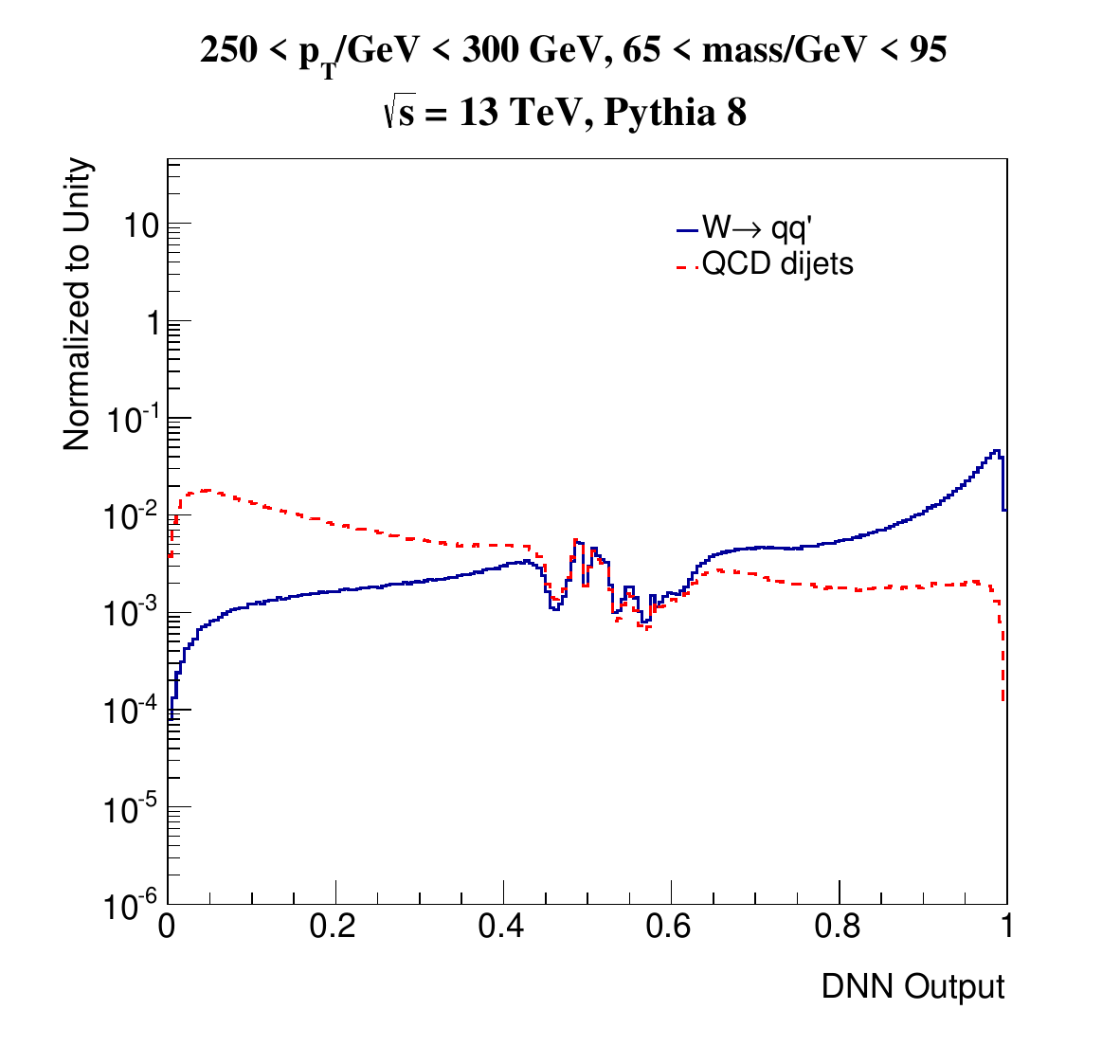}
      \caption{ The marginal distributions of the ConvNet (left) and MaxOut (right) network outputs for signal and background jet images.
      \label{fig:marginal_DNN} }
    \end{center}
\end{figure}

Figure~\ref{fig:marginal_DNN} shows the marginal distributions of the network outputs for signal and background jets.  The MaxOut network has a wavy feature in the distribution near 0.5 where the likelihood ratio is unity.  In that regime, the network cannot differentiate between signal and background and in this particular case results in a non-smooth distribution at the fixed likelihood ratio value.

\begin{figure}[htbp!]
  \begin{center}
        \includegraphics[width=0.45\textwidth]{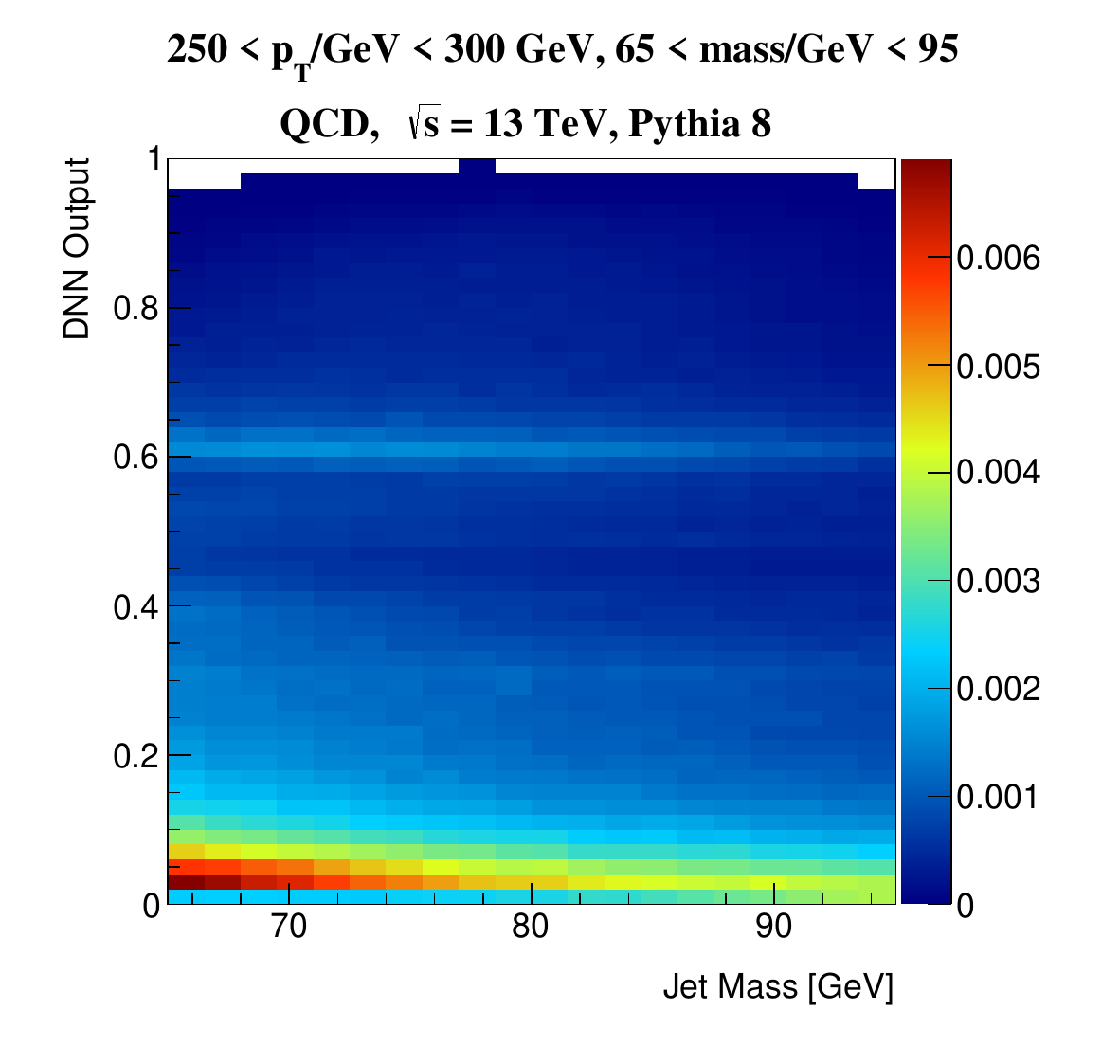} \includegraphics[width=0.45\textwidth]{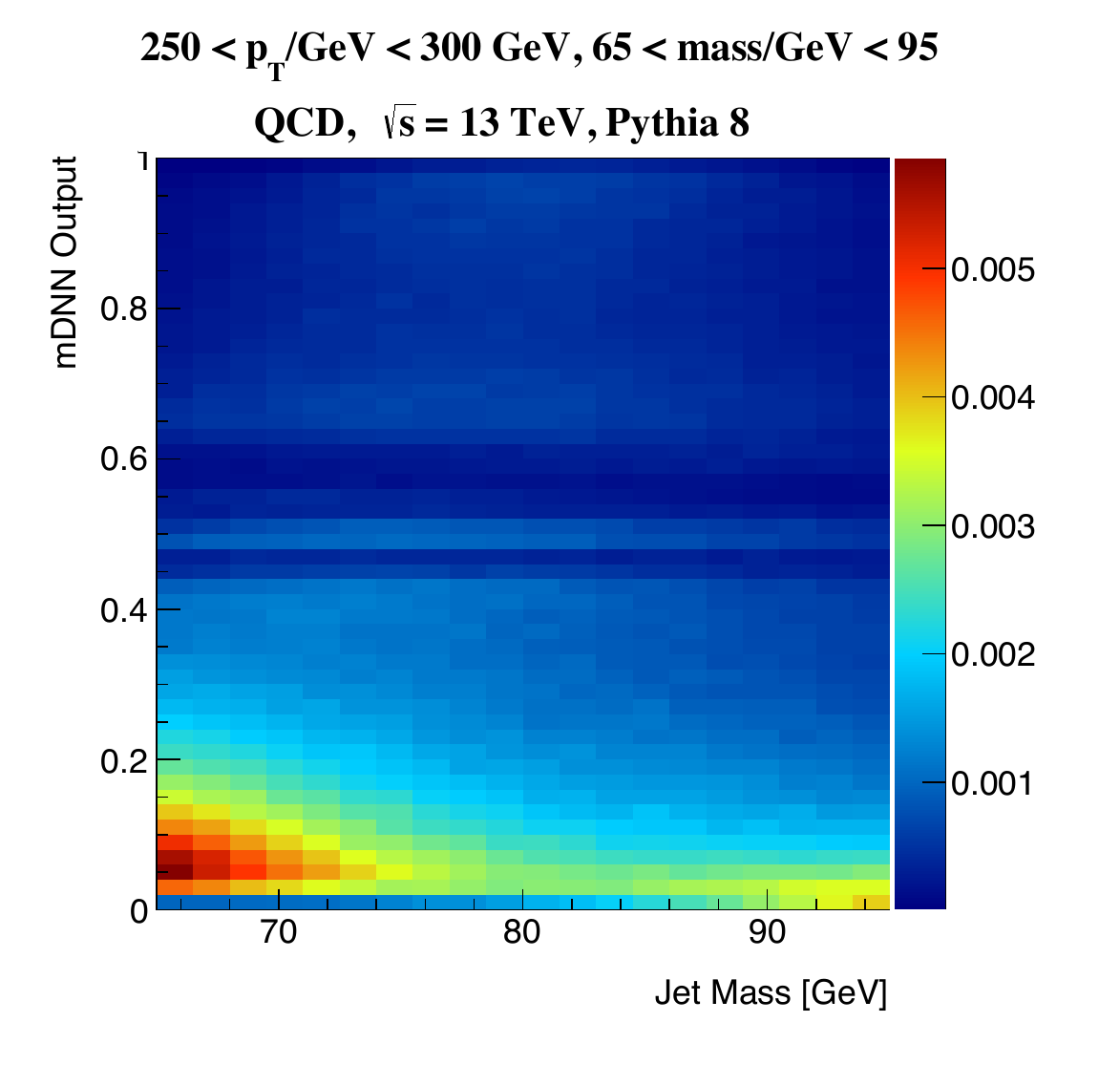}
      \caption{ The joint probability distribution the jet mass and the ConvNet (left) and MaxOut (right) network outputs for the background. 
      \label{fig:mass_DNN} }
    \end{center}
\end{figure}

\begin{figure}[htbp!]
  \begin{center}
        \includegraphics[width=0.45\textwidth]{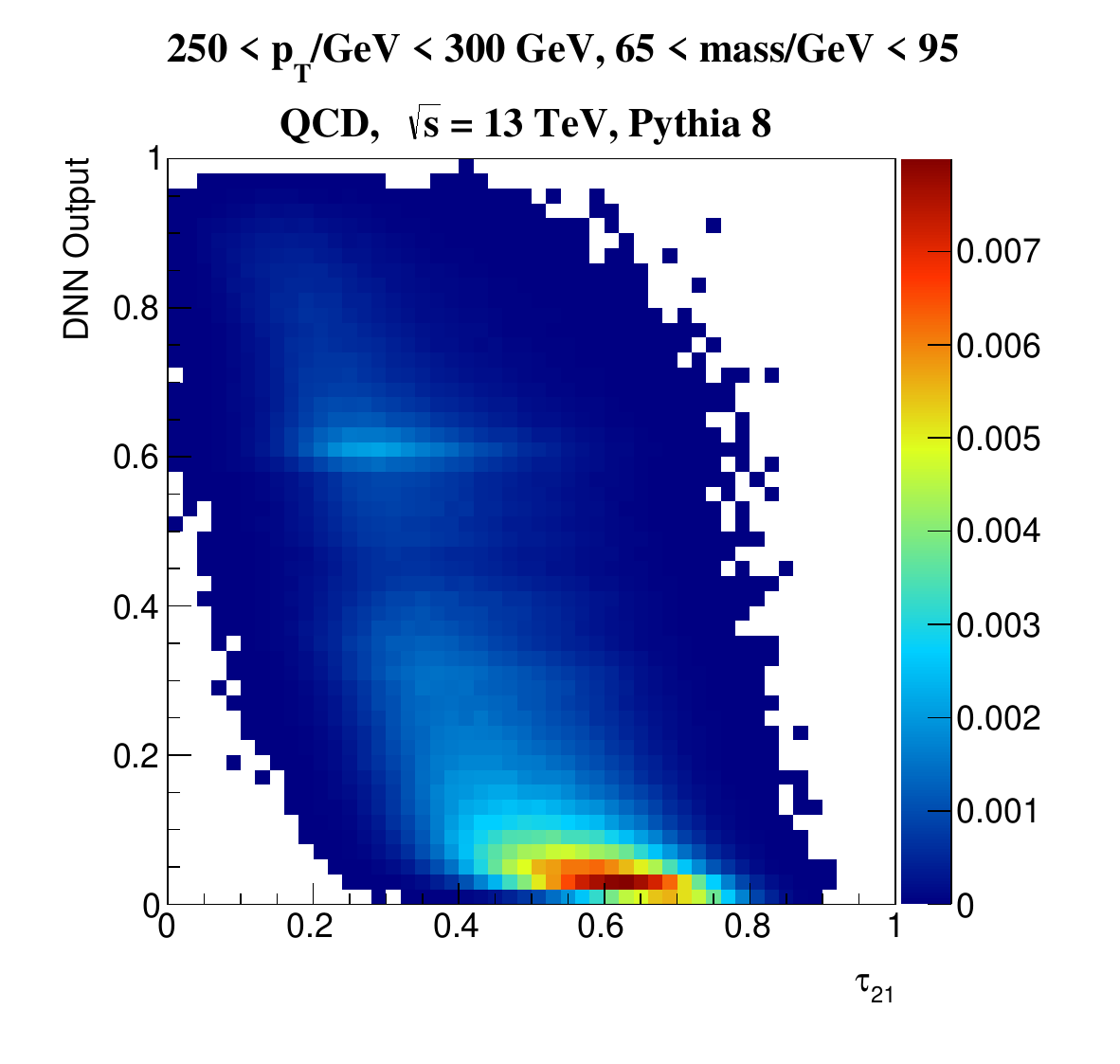} \includegraphics[width=0.45\textwidth]{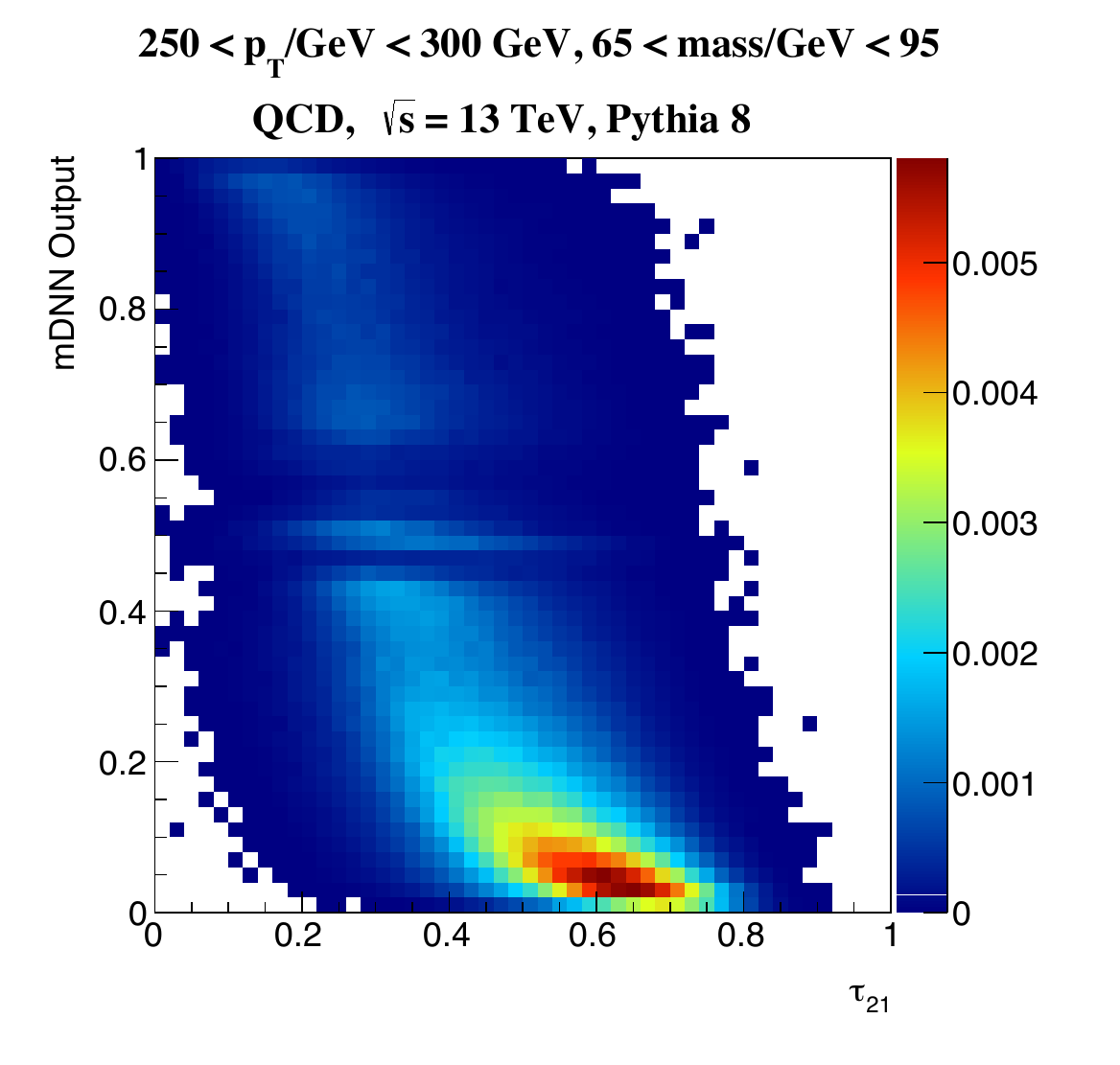}
      \caption{ The joint probability distribution between $\tau_{21}$ and the ConvNet (left) and MaxOut (right) network outputs for the background. 
      \label{fig:tau21_DNN} }
    \end{center}
\end{figure}

\begin{figure}[htbp!]
  \begin{center}
        \includegraphics[width=0.45\textwidth]{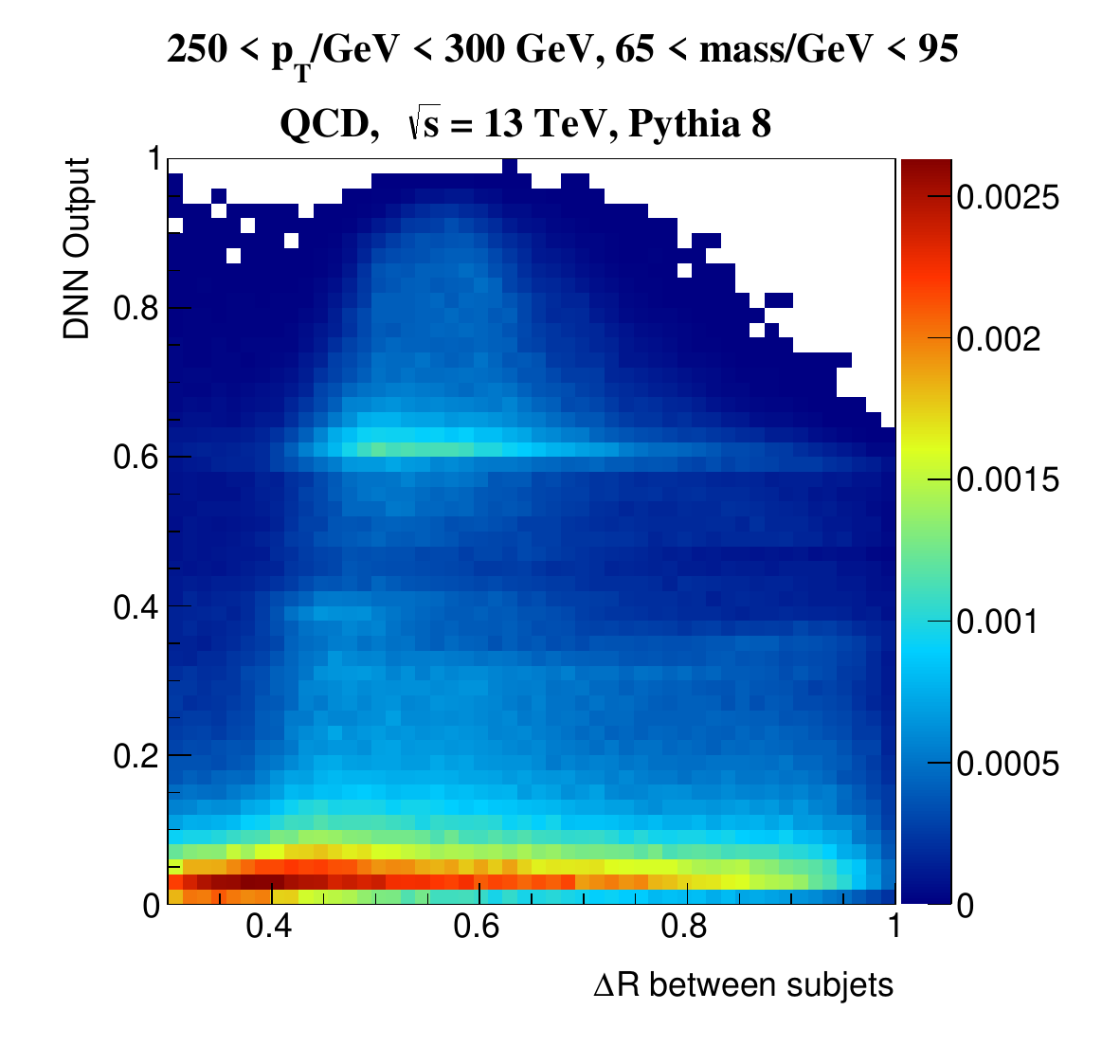} \includegraphics[width=0.45\textwidth]{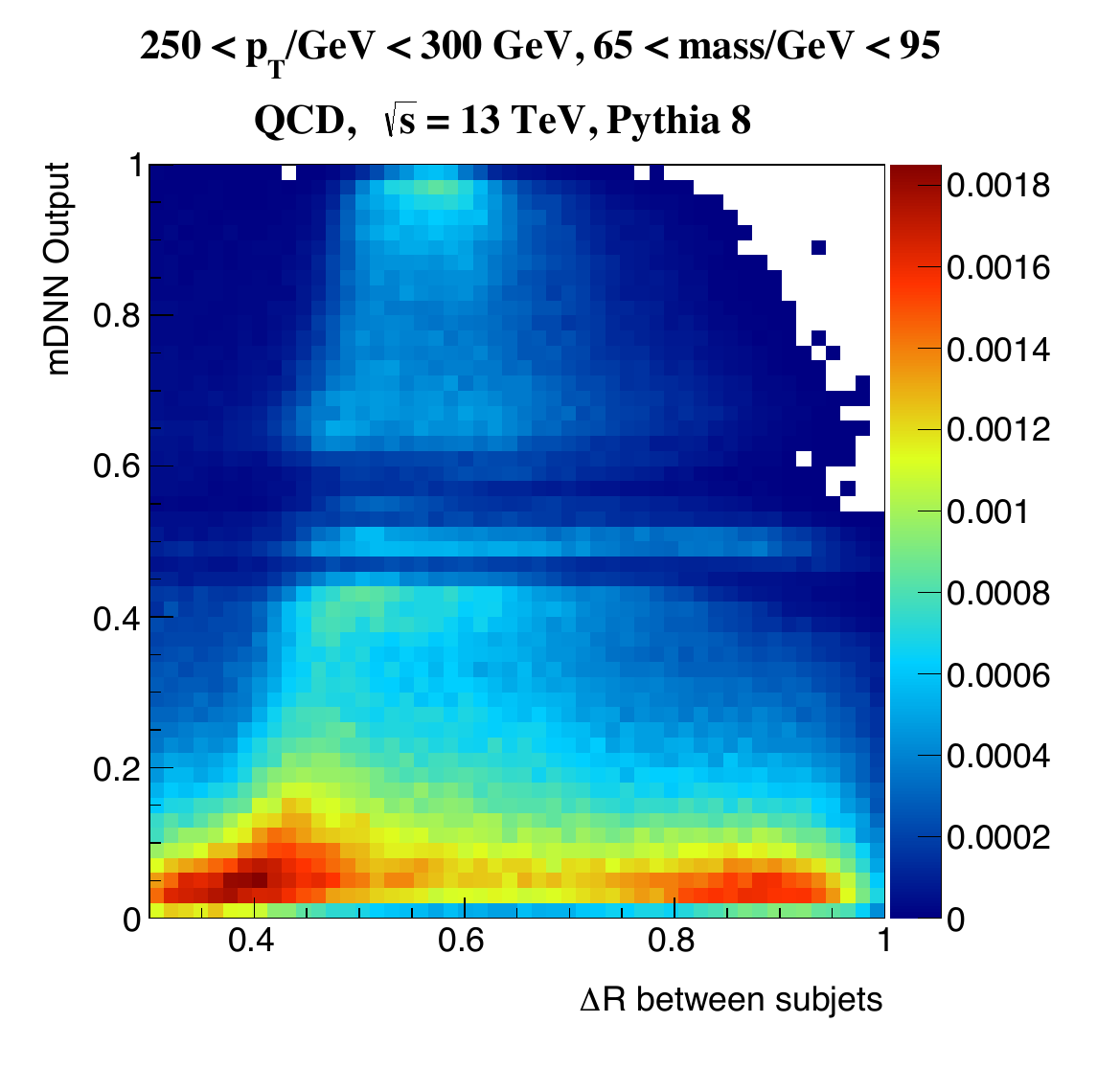}
      \caption{ The joint probability distribution between the $\Delta R$ between subjets and the ConvNet (left) and MaxOut (right) network outputs for the background. 
      \label{fig:dr_DNN} }
    \end{center}
\end{figure}

\begin{figure}[htbp!]
  \begin{center}
        \includegraphics[width=0.45\textwidth]{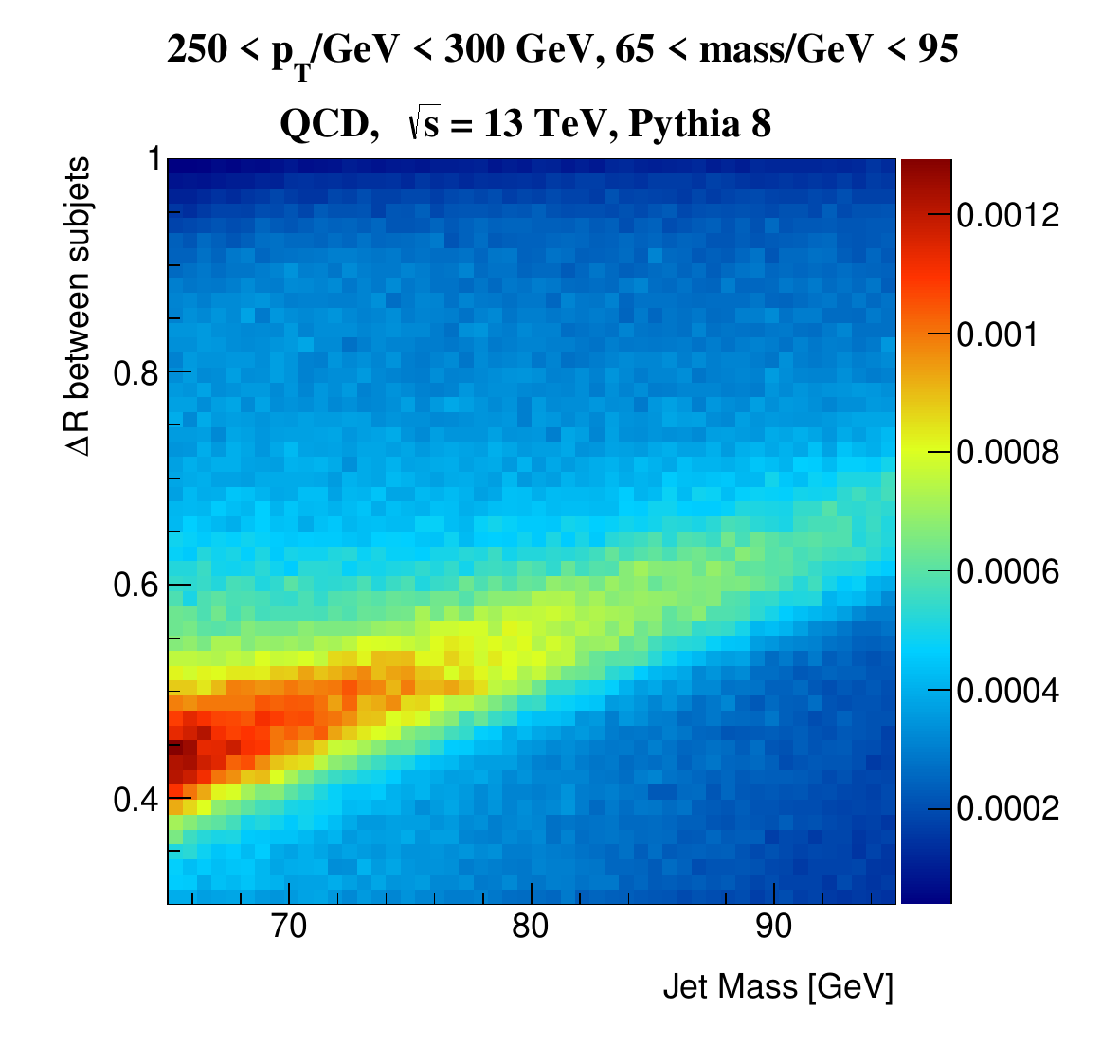} \includegraphics[width=0.45\textwidth]{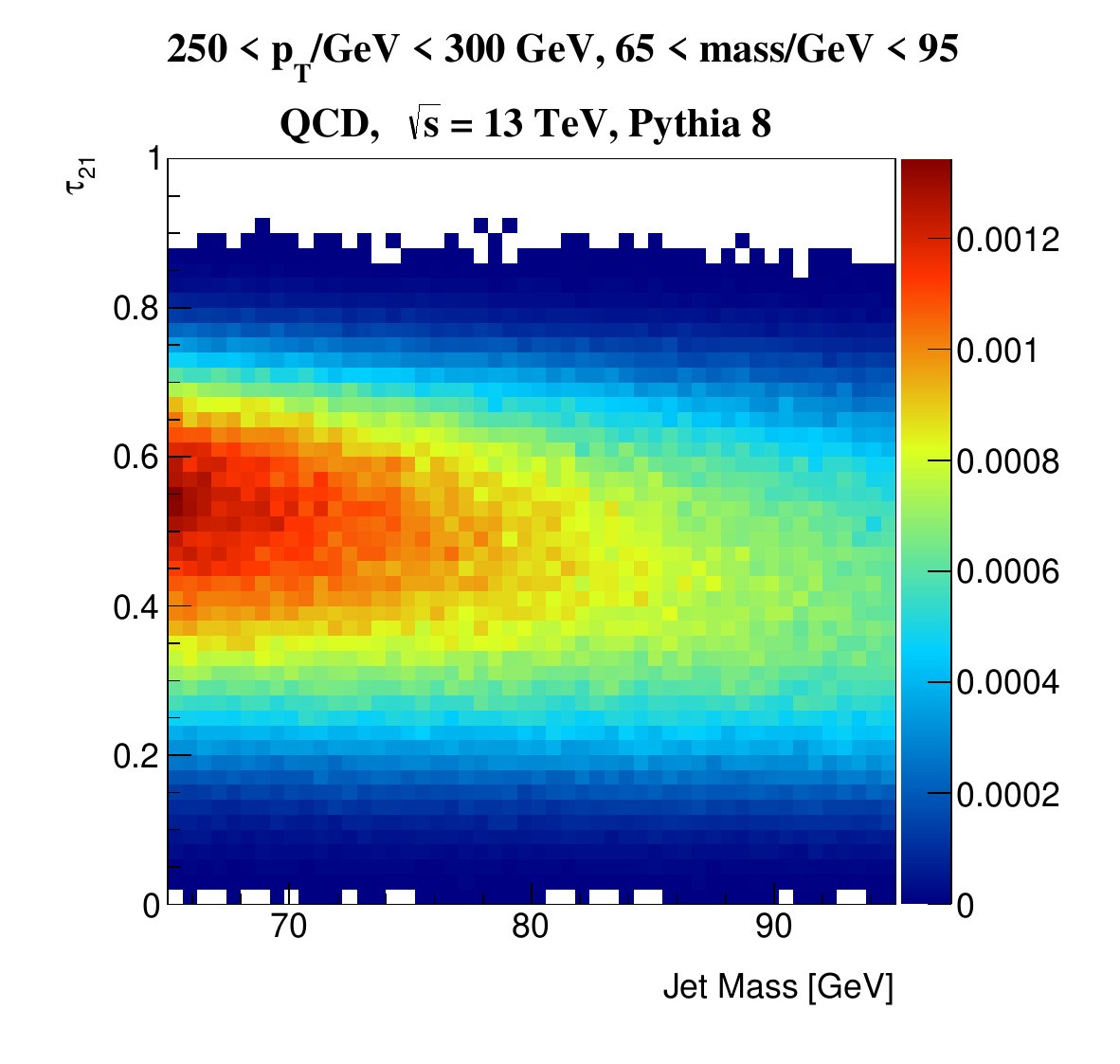}
      \caption{ The joint probability distribution between jet mass and the $\Delta R$ between subjets (left) and $\tau_{21}$ (right) for the background.  
      \label{fig:mass_tau21} }
    \end{center}
\end{figure}

\begin{figure}[htbp!]
  \begin{center}
        \includegraphics[width=0.45\textwidth]{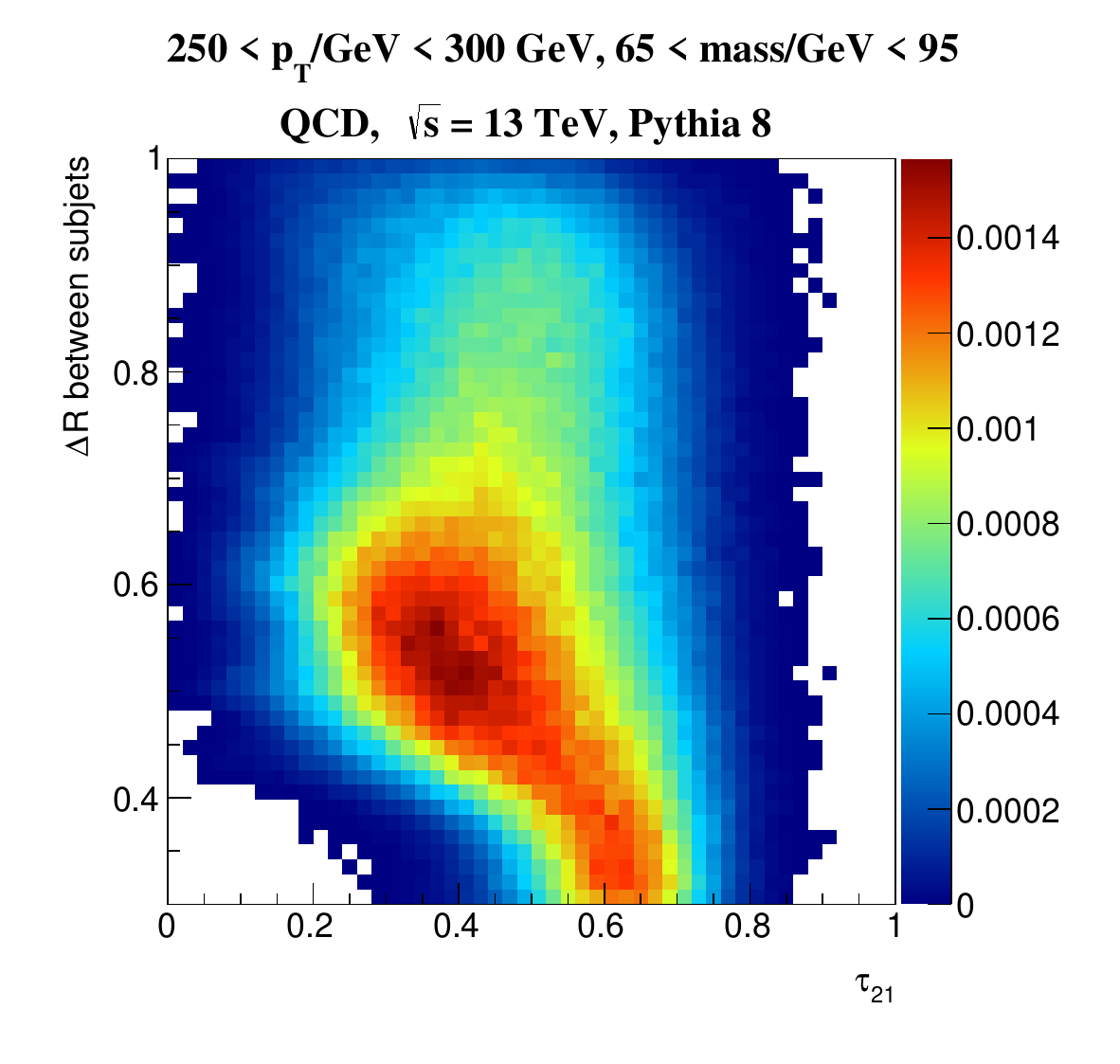}
      \caption{ The joint probability distribution between the $\Delta R$ between subjets and $\tau_{21}$ for the background.      \label{fig:dr_tau21} }
    \end{center}
\end{figure}

The joint distributions of the network with the jet mass, $\tau_{21}$, and the $\Delta R$ between subjets are shown in Fig.~\ref{fig:mass_DNN}, Fig.~\ref{fig:tau21_DNN}, and Fig.~\ref{fig:dr_DNN}, respectively.  The joint distributions between the various combinations of the physics features are shown in Fig.~\ref{fig:mass_tau21} and Fig.~\ref{fig:dr_tau21}.

\clearpage
\newpage

 \bibliography{myrefs}

\end{document}